\documentclass[fleqn,usenatbib]{mnras}
\newcommand{\psj}{PSJ}

\usepackage{newtxtext,newtxmath,comment}
\usepackage[T1]{fontenc}

\DeclareRobustCommand{\VAN}[3]{#2}
\let\VANthebibliography\thebibliography
\def\thebibliography{\DeclareRobustCommand{\VAN}[3]{##3}\VANthebibliography}
\usepackage{graphicx}
\usepackage{amsmath,empheq}
\usepackage{CJKutf8}

\title[HJ Migration]{Constraining Tidal Migration with the Hot Jupiter Population}

\author[Ma et al.]{
\newauthor
Linhao Ma \begin{CJK*}{UTF8}{gbsn}(马林昊)\end{CJK*}$^{1,2}$\thanks{E-mail: linhaoma@princeton.edu},
Yubo Su \begin{CJK*}{UTF8}{gbsn}(苏宇博)\end{CJK*}$^{1,3}$,
Samuel W. Yee$^{4}$\thanks{Heising-Simons Foundation 51 Pegasi b Fellow},
Caleb Lammers$^{1}$,
\newauthor
Eliot Quataert\begin{CJK*}{UTF8}{gbsn} (奎寅略)\end{CJK*}$^{1}$,
and Joshua N. Winn \begin{CJK*}{UTF8}{gbsn}(温乔书)\end{CJK*}$^{1}$
\newauthor
\\
$^{1}$Department of Astrophysical Sciences, Princeton University, 4 Ivy Lane, Princeton, NJ 08544, USA\\
$^{2}$Kavli Institute for Theoretical Physics, University of California, Santa Barbara, CA 93106, USA\\
$^{3}$Canadian Institute for Theoretical Astrophysics, 60 St. George Street, Toronto, ON M5S 3H8, Canada\\
$^{4}$Department of Physics \& Astronomy, University of California Los Angeles, Los Angeles, CA 90095, USA\\
}

\date{}

\pubyear{\the\year{}}

\begin{document}
\label{firstpage}
\pagerange{\pageref{firstpage}--\pageref{lastpage}}
\maketitle

\begin{abstract}
Hot Jupiters with orbital
periods shorter than a few days have probably been affected by tidal
orbital migration. We develop an analytical framework for constraining tidal migration from the present-day hot Jupiter period distribution,
taking into account the
uncertain rate and period distribution of hot Jupiters produced by mechanisms such as high-eccentricity migration or disk-driven migration.
Assuming the tidal migration timescale  is proportional
to $P^{\chi_\tau}$, 
solutions with $\chi_\tau \simeq 3, 1.7,$ and 5.6 are all compatible
with the present-day period distribution.
The $\chi_\tau \simeq 3$ solution 
is consistent with equilibrium tides with
suppression of 
dissipation at short periods,
and implies that newly
circularized hot Jupiters have
periods concentrated
near $3-4$ days, as predicted in some
high-eccentricity migration models.
The $\chi_\tau \simeq 1.7$ solution
is also compatible with the 
$3-4$ day peak
but has
no clear counterpart in existing tidal theories and is more finely tuned.
The $\chi_\tau \simeq 5.6$
solution is compatible with enhanced
short-period equilibrium tidal
dissipation or weakly nonlinear
gravity-wave dissipation,
but requires circularization
at unexpectedly short periods.
Thus, we find the model with $\chi_\tau \simeq 3$ most appealing.
Transit timing of individual
systems and observational
constraints on the rate of
hot Jupiter engulfment provide
additional constraints, which are presently inconclusive but should improve with future data. Improved
measurements of the occurrence of
short-period planets as a function
of planet mass and system age
could also help to sharpen the constraints on tidal migration.
\end{abstract}


\section{Introduction}

The origin of hot Jupiters with
unusually short orbital periods --
a few days or less -- remains uncertain. 
One possibility is that these planets
are delivered onto 
nearly circular orbits with somewhat
longer periods and subsequently migrate
inward as a consequence of the dissipation
of tides raised on their host stars.
However, the dominant mechanism for tidal
dissipation is unknown; different
theories make divergent predictions about
the rate and its dependence on the forcing
period (see, e.g., \citealt{Ogilvie2014}).
Determining that dependence, the ``tidal migration law,'' would therefore
provide a useful empirical test of
theories for tidal dissipation,
with ramifications in other areas
of stellar astrophysics.

A number of observational probes of planetary orbital decay have been considered in the literature \citep{Jackson2009,Pont2009,Teitler2014,Penev2018,Chontos2024}.
Among these probes are the detection of
a shrinking orbital period for the
hot Jupiter WASP-12\,b based on transit
timing observations since 2008  \citep{Hebb+2009, Patra2017,Maciejewski2016,Yee2020}
and constraining upper limits that have been
achieved for a small number of other systems \citep{Patra2017,Adams2024}.
In addition, if tidal migration occurs within
the main-sequence lifetimes of typical host stars,
it should leave an imprint on the present-day
population.
Planets spend less time at periods where
migration is rapid and might tend
to pile up where
migration is slow or where the circularization
process deposits them initially.
This raises the possibility
of inferring the migration law
and initial period distribution
from the present-day period distribution.

Pursuing this goal, \citet{Millholland2025} modeled
the population of hot Jupiters
with an analytic continuity equation
for the distribution of their tidal
decay times, $\tau_d$. They showed
that the short-$\tau_d$ portion of
the distribution approaches a steady
state whose slope depends on the tidal
migration law.
By comparing this
slope with the observed hot Jupiter
population, they found
that the tidal
migration timescale depends only weakly on the period, in a manner consistent
with the theory of tidal resonance
locking \citep{Ma2021}.
They also argued that this
inference is relatively
insensitive to assumptions 
about the rate at which hot Jupiters
are supplied and their initial period distribution.

In this work, we revisit this problem using
a similar continuity-equation approach but with a few important differences.
We work directly with the period
distribution of hot Jupiters
rather than the distribution
of decay times.
Instead of focusing on steady-state solutions, we develop an analytical framework that gives the general time-dependent period distribution, which helps to identify the period range that is most affected by tidal migration. We fit the
analytic model to
a new measurement of the hot-Jupiter period distribution
from Yee et al.\ (in prep.), and
thereby attempt to infer the tidal migration law.

This paper is organized as follows. We describe the analytical problem setup and the governing equations for the migrated hot Jupiter population in Section~\ref{sec:problem_setup}.
We present the analytical solution for the hot Jupiter period distribution in Section \ref{sec:solution_analysis},
with mathematical details provided in Appendices \ref{sec:definition_of_power_law_indices}, \ref{sec:number_density_expression}, and \ref{sec:analysis_details}. We describe the data in \ref{ss:dataset} and the fits to the model in \ref{subsec:data_fitting}. We discuss the possible correspondences
between the model parameters and
theories of tidal dissipation 
in Section \ref{sec:discussion}.
Our results differ from those of \citet{Millholland2025}, for reasons
described in this section.
This section also describes
additional constraints on the tidal
migration law, including a comparison to a newly measured upper limit of the migration rate of WASP-19\,b (Appendix \ref{app:wasp19b}), as well as some possible generalizations of our framework. Finally, we conclude in Section \ref{sec:conclusion}.

\section{Problem Setup}
\label{sec:problem_setup}

We aim to study the long-term tidal migration of hot Jupiters that are already on circular orbits. For these planets, migration can be described solely by the time evolution of their orbital periods $P$. We define the tidal migration timescale
\begin{equation}
\tau_{\rm mig} \equiv \frac{P}{|\dot{P}|}.
\end{equation}
For simplicity, we assume that $\tau_{\rm mig}$ depends only on $P$ and not on any other property of the planet or its host star (see discussion in Section \ref{sec:caveats}). Furthermore, we assume
a power-law dependence:
\begin{equation}
\label{eq:tau_mig}
    {\tau_\mathrm{mig}}=\tau_0\bigg(\frac{P}{P_0}\bigg)^{\chi_\tau}\,,
\end{equation}
where $\tau_0$ is the tidal migration timescale at the fiducial period $P_0$. This parameterization can describe a number of different theories of tidal migration with different choices of $\chi_\tau\geq0$ and $\tau_0$, as we will summarize in Section \ref{sec:migration_theories}.

We consider an ensemble of systems and describe the planet population by the period-specific number density $n=n(P,t)$, defined
such that $n(P,t)\,dP$ is the number of planets
with periods between $P$ and $P+dP$ at time $t$. The population is supplied by planets that are delivered onto short-period, nearly circular orbits through processes such as disk-driven migration
or high-eccentricity migration.
The supplying mechanism
is described by the source function $S(P, t)$, defined
such that
$S(P, t)\,dP\,dt$ is
the number of hot Jupiters
added to the population with periods
between $P$ and $P+dP$ during the time
interval $dt$.
Thus, $t$ is the time over which the population has been
continuously supplied by the
source function.

For simplicity, we assume the source
function is time-independent, i.e.,  $S(P,t)=S(P)$. 
We do not include a sink function
because although hot
Jupiters may ultimately be destroyed
by tidal disruption, this
occurs at extremely short periods ($P\lesssim 0.5\,\mathrm{days}$) where the current data provide only
upper limits (see Figure \ref{fig:fit}).
Nevertheless, we derive an inferred destruction rate from our results in Section \ref{sec:destruction}, and our formalism can be generalized to include a sink function as discussed in Section \ref{sec:caveats}.

With these definitions, $\dot{P} n$ is the flux of planets
through period space due to tidal migration,
and the evolution of the hot Jupiter population is governed by a one-dimensional continuity equation:
\begin{equation}
\label{eq:continuity}
    \frac{\partial n}{\partial t}+\frac{\partial(\dot{P} n)}{\partial P}=S\,.
\end{equation}
The solution depends on both the migration-rate parameterization and the form of the source function.

\section{Solution and Analysis}
\label{sec:solution_analysis}

For a given initial condition,
we can solve Equation \ref{eq:continuity} using the Green's function method \citep{GreenEssay}. Assuming there are no circularized planets at $t=0$, we show in Appendix \ref{sec:number_density_expression} that the formal solution to Equation \ref{eq:continuity} is
\begin{equation}
\label{eq:formal_solution}
    n(P,t)=\frac{\tau_\mathrm{mig}}{P}\int_P^{P_t}S(P')dP'\,.
\end{equation}
Here, $P_t$ is the period for which a planet requires a time $t$
to migrate to a period $P$, and is given by
\begin{empheq}[left={P_t(P,t)=\empheqlbrace}]{equation}
\label{eq:Pt}
  \begin{aligned}
      & P(1+\chi_\tau t/\tau_\mathrm{mig})^{1/\chi_\tau}&&,\;\chi_\tau>0,\\
      &P\exp(t/\tau_\mathrm{mig})&&,\;\chi_\tau=0\,.
  \end{aligned}
\end{empheq}
Thus, the integral in Equation~\ref{eq:formal_solution}
gives the total rate at which new planets are supplied onto
trajectories that can reach $P$ within time $t$.
The factor $\tau_{\rm mig}/P = |\dot{P}|^{-1}$ converts
this rate into a period-space number density.

In general, $n(P,t)$ depends on both the tidal migration law $\tau_\mathrm{mig}(P)$ and the source function $S(P)$. Nevertheless, in certain cases, the shape of $n(P,t)$ is only weakly dependent on the source function, allowing
the migration law to be constrained directly from the observed
population. We discuss these cases below.

\subsection{$\chi_\tau=0$, resonance locking}
\label{sec:res_lock}

A special case of interest is when a planet migrates through tidal resonance locking with its host star's oscillation modes (\citealt{Ma2021}, see Section \ref{sec:migration_theories}
for further explanation). This theory predicts $\chi_\tau=0$
and $\tau_{\rm mig} = \tau_0$, i.e., the tidal migration timescale is independent of the orbital period. In this case,
\begin{equation}
    n(P,t)=\frac{\tau_0}{P}\int^{P\exp(t/\tau_0)}_PS(P')dP'\,.
\end{equation}
If the source function is a power-law over this interval,
$S\propto P^{\chi_s}$,
then the migrating population maintains the same power-law dependence,
$n\propto P^{\chi_s}$ (see Appendix~\ref{sec:analysis_rl}).
This occurs because for $\chi_\tau = 0$, all planets experience
a steady decrease in $\ln P$. Migration simply translates
the population in $\ln P$ without changing the shape of its period distribution.

\subsection{$\chi_\tau\sim \mathcal{O}(1)>0$, period-dependent migration}

Theories other than resonance locking typically predict an order-of-unity $\chi_\tau>0$, i.e., a migration timescale that increases with orbital
period (see Section \ref{sec:migration_theories}). We generally call such scenarios ``period-dependent'' migration, meaning that the migration rate is a strong function of orbital period. 
In this case, it is useful to define
\begin{equation}
\label{eq:Pmig}
P_\mathrm{mig}(t) \equiv P_0\left(\frac{\chi_\tau t}{\tau_0}\right)^{1/\chi_\tau}\,.
\end{equation}
Under this definition, Equation \ref{eq:tau_mig} reduces to:
\begin{equation}
\label{eq:tau_mig_P_mig}
    {\tau_\mathrm{mig}}\equiv-\frac{P}{\dot{P}}=\chi_\tau t\,\bigg(\frac{P}{P_\mathrm{mig}}\bigg)^{\chi_\tau}\,.
\end{equation}
Thus, $P_\mathrm{mig}$ is the characteristic orbital period below which tidal migration has substantially modified the population
over time $t$. 
Since it is difficult to predict the form of the source function $S$ from first principles, we explore two
cases that might be of physical interest.

\begin{figure*}
    \centering
\includegraphics[width=\linewidth]{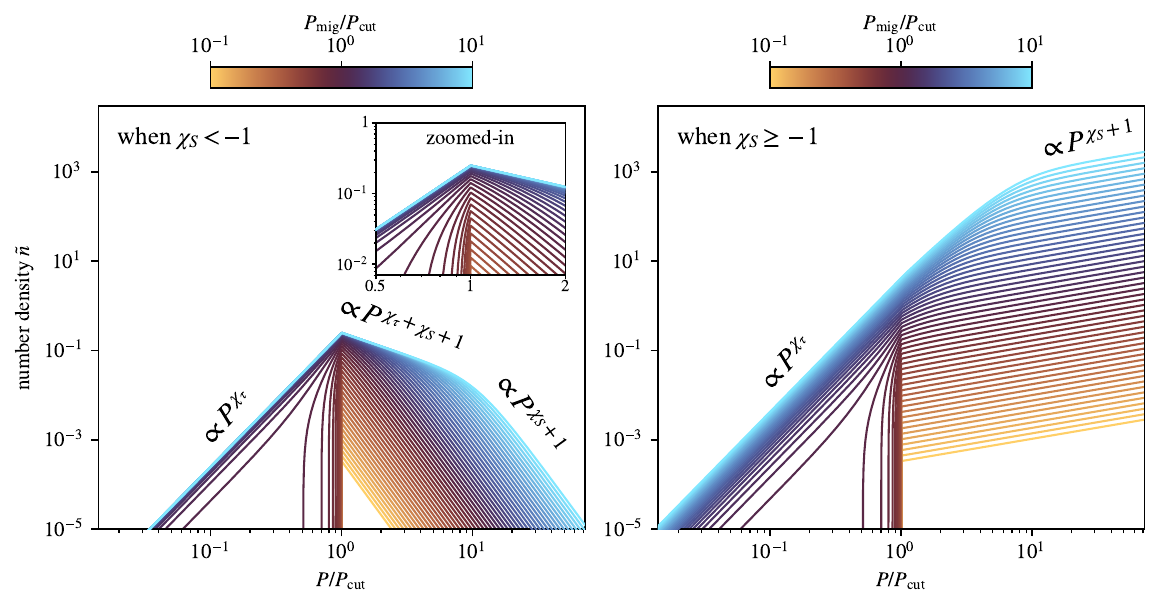}
    \caption{Numerical solutions of the log-space planet number density $\tilde{n}=Pn(P,t)$ for different values of $P_\mathrm{mig}/P_\mathrm{cut}$, assuming a tidal migration timescale proportional to $P^{\chi_\tau}$ and a power-law source function proportional to $P^{\chi_S}$ with a lower truncation at $P_\mathrm{cut}$ (Equation \ref{eq:powerlaw_source}). $P_\mathrm{mig}$ is the orbital period  below which tidal migration has substantially modified the population at time $t$.   When expressed in terms of $P/P_\mathrm{cut}$, the shape of $\tilde n$ is solely determined by $P_\mathrm{mig}/P_\mathrm{cut}$, which increases with time. The left panel shows the case $\chi_S<-1$, with an inset panel showing more detail where  $0.5<P/P_\mathrm{cut}<2$.
    The right panel shows the case  $\chi_S\geq-1$. When $P_\mathrm{mig}\ll P_\mathrm{cut}$ (gold curves), hot Jupiters are generated beyond the period where migration is significant, hence the number density resembles the source function integrated over time. When $P_\mathrm{mig}\approx P_\mathrm{cut}$ (dark brown curves, more easily seen in the inset panel), some planets that are
    sourced just above the truncation period $P_\mathrm{cut}$
    migrate to shorter periods, creating a steep and non-power-law drop in number density toward shorter periods. When $P_\mathrm{mig}\gg P_\mathrm{cut}$ (blue curves), planets migrate significantly from a wider range of initial periods. The number density develops into a broken power-law function with two or three zones,
    depending on whether $\chi_S<-1$ or $\chi_S\gtrsim-1$. The power-law indices are generally linear combinations of $\chi_\tau$ and $\chi_S$, as labeled near the curves and derived in Appendix \ref{appendix:truncated_power_law}.}
    \label{fig:power-law-s}
\end{figure*}

\subsubsection{Truncated power-law source function}
In the first case, we assume the source function is a power law, with a lower truncation at $P_\mathrm{cut}$:
\begin{empheq}[left={S(P)=\empheqlbrace}]{equation}
\label{eq:powerlaw_source}
  \begin{aligned}
      & 0&&,\; P<P_\mathrm{cut},\\
      &S_0\bigg(\frac{P}{P_\mathrm{cut}}\bigg)^{\chi_S}&&,\; P\geq P_\mathrm{cut}\,.
  \end{aligned}
\end{empheq}
The truncation period may represent the magnetic truncation radius in disk-driven migration or the characteristic circularization period in high-eccentricity migration (see, e.g., \citealt{Wu2018}). In principle, there also exists an upper truncation
period, but in realistic
theories, it is likely beyond $P_\mathrm{mig}$, making it uninteresting in a study of short-period tidal migration.

The solution of $n(P,t)$ with the specified source function is\footnote{As written, these expressions
assume $\chi_S\neq -1$,
but the solution for
$\chi_S=-1$ can be obtained
by taking the ${\chi_S \to -1}$
limits.}:
\begin{empheq}[left={n(P,t)=\empheqlbrace}]{equation}
\label{eq:n_solution_power_law_chis_neq_minus_1}
  \begin{aligned}
      &\,0&&,\; \mathcal{P}_t<1,\\
      &\,n_0\,\mathcal{P}^{\chi_\tau-1}\left(\mathcal{P}_t^{\chi_S+1}-1\right)&&,\; \mathcal{P}< 1\leq \mathcal{P}_t,\\
      &\,n_0\,\mathcal{P}^{\chi_\tau-1}\left(\mathcal{P}_t^{\chi_S+1}-\mathcal{P}^{\chi_S+1}\right)&&,\; \mathcal{P}\geq1\,,
  \end{aligned}
\end{empheq}
where 
$\mathcal{P}\equiv P/P_\mathrm{cut}$, $\mathcal{P}_t\equiv P_t/P_\mathrm{cut}$,
and 
\begin{equation}
n_0\equiv S_0\tau_0\,\frac{(P_\mathrm{cut}/P_0)^{\chi_\tau}}{\chi_S+1}
\end{equation}
is a common normalization factor independent of $P$ and $t$. From the definitions of $P_t$ and $P_\mathrm{mig}$ (Equations \ref{eq:Pt} and \ref{eq:Pmig}), one can show that:
\begin{equation}
    \mathcal{P}_t=(\mathcal{P}^{\chi_\tau}+\mathcal{P}_\mathrm{mig}^{\chi_\tau})^{1/\chi_\tau}\,,
\end{equation}
where $\mathcal{P}_\mathrm{mig}\equiv P_\mathrm{mig}/P_\mathrm{cut}$. Therefore, the dependence of $n(P,t)$ on $\mathcal{P}$ (hence $P$ for a fixed $P_\mathrm{cut}$) is solely determined by the parameter $\mathcal{P}_\mathrm{mig}$,
up to a common normalization factor $n_0$.

For visualizing the results and comparing them to data, it will be useful to examine the number density of planets per unit $\ln P$ rather than $P$. We denote this $\ln P$-space number density by ${\tilde {n}}$,
which obeys
\begin{equation}
\label{eq:tilde_n}
    \tilde {n}(P,t)=n(P,t)\frac{dP}{d\ln P}=Pn(P,t)\,.
\end{equation}

In Figure \ref{fig:power-law-s}, we show the numerical solutions of ${\tilde n}(P,t)$ as a function of $\mathcal{P}\equiv P/P_\mathrm{cut}$ for different values of $\mathcal{P}_\mathrm{mig}\equiv P_\mathrm{mig}/P_\mathrm{cut}$, with the same normalization factor and a fixed choice of $\chi_S$ and $\chi_\tau$. As $P_\mathrm{cut}$ is independent of time, $P_\mathrm{mig}/P_\mathrm{cut}$ increases as time evolves, and the overall number of planets at long periods increases as more planets are supplied. The exact shape of the curve also changes and depends on whether $\chi_S$ is greater than $-1$ or not.

First, consider $\chi_S<-1$ (left panel).
At early times
($P_\mathrm{mig}<P_\mathrm{cut}$), the number density resembles the source function (gold lines). This is because all newly supplied planets have periods longer than $P_\mathrm{mig}$ and have not had enough time to migrate significantly.
As time progresses and $P_\mathrm{mig}$ approaches $P_\mathrm{cut}$, the planets with periods near
$P_\mathrm{cut}$ experience significant
period decay, creating a population of planets with $P<P_\mathrm{cut}$ (dark brown lines, most easily seen in the zoomed-in panel in Figure \ref{fig:power-law-s}). At late times,
$P_\mathrm{mig} > P_\mathrm{cut}$, $\tilde n$ approaches a three-zone power-law function (blue lines).
Appendix \ref{sec:analysis_freq_dependent_migration}
works out this limiting case analytically and derives
the power-law indices. Here, we summarize the results:
\begin{itemize}

    \item When $P<P_\mathrm{cut}$, 
    $\tilde n \propto P^{\chi_\tau}$.
    Planets with periods in this range are not
    supplied by the source function; they occur
    only as a consequence of tidal migration
    from longer periods. Thus, the power-law index in
    this regime depends chiefly on the tidal migration law, with only a weak dependence on the source function.

   \item When $P_\mathrm{cut}<P<P_\mathrm{mig}$,
   $\tilde n\propto P^{\chi_\tau+\chi_S+1}$. 
   The population is a mixture
   of newly supplied planets and tidally migrated planets,
   and the power-law index depends on both $\chi_\tau$ and $\chi_S$.

   \item  When $P>P_\mathrm{mig}$,
   $\tilde n\propto P^{\chi_S+1}$. This is the slow-migration limit in which the number density simply
   reflects the source function.

\end{itemize}

Next, consider $\chi_S\geq-1$ (right panel).
The results are similar to the previous case
for $P<P_\mathrm{cut}$ and $P>P_\mathrm{mig}$. The only difference is that when $P_\mathrm{cut}<P<P_\mathrm{mig}$, we have $\tilde n\propto P^{
\chi_\tau}$, which is similar to the case of $P<P_\mathrm{cut}$. This makes the number density when $P_\mathrm{mig}>P_\mathrm{cut}$ a two-zone power-law function instead of the three-zone function as in the case of $\chi_S<-1$, as seen in Figure \ref{fig:power-law-s}.

Generally, we can see that aside from the case where no planets have migrated significantly, the number density at shorter periods is mostly sensitive to the tidal migration law (parameterized by $\chi_\tau$), while at longer periods, it is more sensitive to the source function (parameterized by $\chi_S$). Therefore, if the data appear to be a two-zone or three-zone power-law function, it may become possible to distinguish the contributions from the source function and the migrated population by fitting the short- and
long-period ends of the distribution, subject
to the assumption of a power-law
source function.

\subsubsection{Gaussian source function}
In the second case, we assume the source function is a Gaussian function
of the semi-major axis $a$, with mean $\mu_a$ and standard deviation $\sigma_a$:
\begin{equation}
\label{eq:gaussian_source}
    S_a(a)=\frac{S_0}{\sqrt{2\pi \sigma_a^2}}\exp\left(-\frac{(a-\mu_a)^2}{2\sigma_a^2}\right)\,.
\end{equation}
The $P$-space source function is then related to $S_a$ as $S(P)=2aS_a(a)/(3P)$ by applying Kepler's Third Law. This form of the source function is inspired by, e.g., in-situ disk formation of hot Jupiters, where planets are generated over a narrow range of initial semi-major axes. 

Appendix \ref{app:gaussian_source} gives the solution for $\tilde{n}(P,t)$
in this case. To understand
the solution, it is useful
to consider the behavior at very short periods, and then the effect
of gradually increasing the period.
When
$P\ll P_\mathrm{mig}$ and $a\ll \mu_a$, 
the solution has the
form $\tilde n\propto P^{\chi_\tau}$, similar to the case of the power-law source function at short periods.
As $P$ and the corresponding
$a$ are increased,
if $P$ reaches $P_\mathrm{mig}$
 while $a$ is still smaller than $\mu_a$, the planets are in the slow-migration limit, and the original Gaussian source function
is recovered. If $P$ is still less than $P_\mathrm{mig}$ when $a$ reaches $\mu_a$, the number density drops sharply and roughly scales as $\tilde n \propto P^{\chi_\tau-\mathcal{O}(a/\sigma_a)}$. Figure \ref{fig:gaussian_source} shows some illustrative numerical solutions
consistent with our analysis.

In summary, when the source function is Gaussian over the semi-major axis, we expect the number density at short periods to be sensitive to the migration rate only, characterized by $\chi_\tau$. At longer periods, the Gaussian source is recognizable by its exponentially decaying tail because no planets are generated far from the mean of the Gaussian.

\section{Comparing Theory to Data} \label{sec:data_comparison}

In this section, we fit our analytical model to the observed, bias-corrected population of hot Jupiters.

\begin{figure*}
    \centering
    \includegraphics[width=\textwidth]{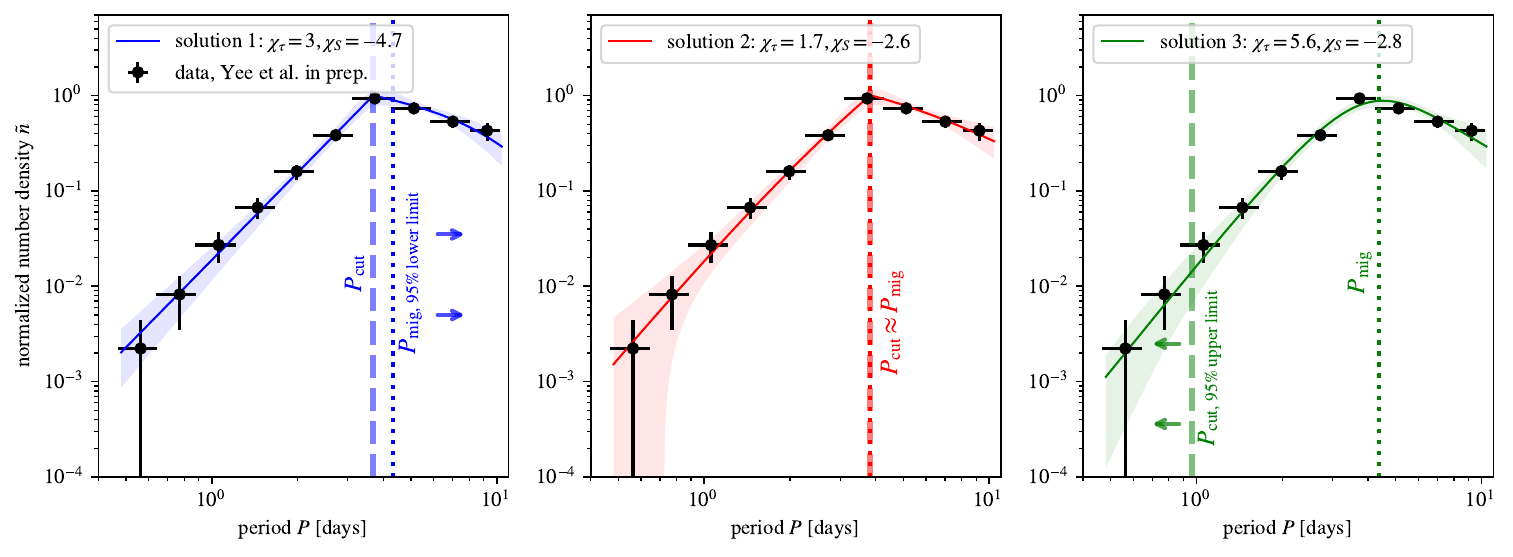}
    \caption{
        Comparison of the observed period distribution of hot Jupiters to
        model predictions. The model
        has two components: a tidal migration timescale $\propto P^{\chi_\tau}$, and a source function  $\propto P^{\chi_S}$ with a lower truncation period $P_\mathrm{cut}$ (Equation \ref{eq:powerlaw_source}).
        The black points are the bias-corrected hot Jupiter occurrence rates binned in $\ln P$, with bin widths shown by the horizontal error bars.
    The panels depict three solutions that provide good fits to the data: the colored curves show the calculated number density using the maximum-likelihood parameters, and the shaded regions show the 95\% credible intervals (Figure \ref{fig:corners}). Solution 1 features $\chi_\tau\approx 3$, similar to the value of $10/3$ predicted by the theory of ``fast tides suppression.'' Solution 2 has $\chi_\tau\approx 1.7$, which is not consistent with any known tidal theory. Solution 3 has $\chi_\tau\approx 5.6$, consistent with the  prediction of 5.7 from ``fast tides enhancement'' or the range
        6--7 predicted by weakly nonlinear wave dissipation. However, solution 3 requires the source function to peak at $P\lesssim 1\,\mathrm{day}$, shorter than expected in standard disk-driven migration or high-eccentricity migration.
    }
    \label{fig:fit}
\end{figure*}

\subsection{Data Selection}\label{ss:dataset}

We obtained the occurrence rate of hot Jupiters as a function of orbital period from a newly-constructed sample of planets from NASA's \textit{TESS} mission \citep{Ricker2015}.
The precise construction and calculations will be described in a forthcoming paper (Yee et al., in prep.), but we summarize the procedure here.
Because hot Jupiters are intrinsically rare, a large number of stars must be searched in order to construct a sample sufficiently large for statistical study.
We used a magnitude-limited sample of 919{,}600 FGK stars ($0.7\,M_\odot \leq M_\mathrm{star} \leq 1.2 \,M_\odot$) brighter than 12.5~mag in the \textit{Gaia} $G$ band.
These stars were selected based on their positions in the extinction-corrected color-magnitude diagram \citep{Yee2021b} and the availability
of TESS data. To maximize the completeness
of this sample, no other selection criteria were imposed.

The planet sample comprises 368 hot Jupiters orbiting these stars for which transits
were detected in the \textit{TESS} data and the implied planet radius is in the range $8\,R_\oplus < R_p < 24\,R_\oplus$.
This includes both planets that were newly discovered using \textit{TESS} data, as well as previously known planets found by earlier ground-based transit searches.
While \textit{TESS} detected many other transiting planet candidates around the magnitude-limited sample of stars, extensive ground-based follow-up allowed us to weed out the false positives, leaving us a clean sample of only confirmed planets.

We modeled the hot Jupiter occurrence rate as an inhomogeneous Poisson point process \citep[e.g.,][]{Tabachik2002,Cumming2008,Foreman-Mackey2014a,Zhu2021}. 
We divided the parameter space into bins, logarithmically in orbital period between $0.3 < P < 10$~days and planet radii $8\,R_\oplus < R_p < 24\,R_\oplus$.
In each bin $j$, we calculated
the effective number of stars searched, $N_{\star,j}^{\mathrm{eff}}$, by summing the product of the geometric transit probability and the detection probability determined by injection-recovery tests over all the stars in the sample.

We found that \textit{TESS}'s detection
sensitivity for transiting hot Jupiters across the parameter space of interest is very high, with a mean detection probability of 97\% over all stars, planet radii, and orbital periods.
For the present work, we are concerned with the orbital period distribution, so we marginalized
the combined detection efficiency over planet
radius at fixed period.
Given $N_{p,j}$ detections in each bin, we computed the posterior probability for the intrinsic planet frequency:
\begin{equation}
    {\rm prob}(\bar{n}_{p,j} | N_{p,j}, N_{\star,j}^{\mathrm{eff}}) \propto \mathcal{L}(N_{p,j} | \bar{n}_{p,j}, N_{\star,j}^{\mathrm{eff}})\, \pi(\bar{n}_{p,j}),
\end{equation}
where $\mathcal{L}$ is a Poisson likelihood
with $\bar{n}_{p,j}\, N_{\star,j}^{\mathrm{eff}}$ as
the expected number of detections.
We adopted a flat non-negative prior $\pi(\bar{n}_{p,j})$, equivalent to
a conjugate
Gamma prior with shape parameter $\alpha=1$ and rate parameter $\beta\rightarrow 0$. The resulting
posterior is
\begin{equation}
    \bar{n}_{p,j} \sim \Gamma(\alpha = N_{p,j} + 1,~ \beta = N_{\star,j}^{\mathrm{eff}}).
\end{equation}
We used the median and 68\% credible interval of this posterior in our subsequent comparisons.

\subsection{Data Fitting}
\label{subsec:data_fitting}

As shown in Figure \ref{fig:fit}, the observed log-space hot Jupiter occurrence rate appears to be a two-zone power-law function with a peak at $3-4$ days (or $a\sim0.05\,\mathrm{AU}$
for a solar-mass star). Comparing this to our numerical solutions in Figure \ref{fig:power-law-s}, we
can immediately exclude the case of $\chi_S>-1$, which predicts an increasing number density at long periods.

As there is still a significant number of hot Jupiters at $P\sim10\,\mathrm{days}$ (or $a\sim0.1\,\mathrm{AU}$ assuming $1\,M_\odot$ hosts), the data also disfavor a narrow Gaussian source function. To match the data, the Gaussian requires a broad width ($\sigma > 0.05\,\mathrm{AU}$), which would be similar to a flat power-law source function at the periods of interest. This led us to adopt a truncated power-law source function with $\chi_S<-1$.

By comparing the data with the numerical solution in the left panel of Figure \ref{fig:power-law-s}, we can see that there are three classes of numerical solutions that may fit the observed two-zone power-law function shown, as we summarize below:

\begin{enumerate}

    \item Solution 1: the turnover at $P\simeq 3\,\mathrm{days}$ corresponds to $P\simeq P_\mathrm{cut}$ and $P_\mathrm{mig}\gg P_\mathrm{cut}$ (blue lines). Below and above $P_\mathrm{cut}$, the number density scales as $\tilde{n}\propto P^{\chi_\tau}$ and $\tilde{n}\propto P^{\chi_\tau+\chi_S+1}$, respectively, giving two different power laws. The absence of a third power-law at long periods (corresponding to $P>P_\mathrm{mig}$) in the data prevents a clear determination of $P_\mathrm{mig}$.
    
    \item Solution 2: the turnover at $P\simeq 3\,\mathrm{days}$ is identified with both
    $P_\mathrm{cut}$ and 
    $P_\mathrm{mig}\approx
    P_\mathrm{cut}$ (dark brown lines). Below $P_\mathrm{cut}$, the number density is not strictly
    a power law, though it may appear similar to a power law over the
    range of periods spanned by the data. 
    Although solution 2 connects
    continuously to solution 1 as
    $P_\mathrm{mig}/ P_\mathrm{cut}$ increases, we regard it as distinct because it corresponds to a qualitatively different regime.
    In solution 1, planets have been
    migrating across $P_{\rm cut}$ long
    enough for the $P<P_{\rm cut}$ population
    to approach
    the asymptotic limit $\tilde{n}\propto P^{\chi_\tau}$.
    In solution 2, by contrast,
    tidal migration has only recently
    begun to populate periods below $P_{\rm cut}$,
    and the distribution
    is transitional rather than asymptotic.
    
    \item Solution 3: the turnover at $P\simeq 3\,\mathrm{days}$ corresponds to $P\simeq P_\mathrm{mig}$ and $P_\mathrm{mig}\gg P_\mathrm{cut}$ (blue lines). Below and above $P_\mathrm{mig}$, the number density scales as $\tilde{n}\propto P^{\chi_\tau+\chi_S+1}$ and $\tilde{n}\propto P^{\chi_S+1}$, respectively, giving two different power laws. In this solution,
    the absence of a third power-law at short periods 
    in the data implies
    that $P_\mathrm{cut}$ is shorter than about 1~day, the shortest
    period for which the data
    are constraining.
\end{enumerate}

We hence fit the data with the numerical solutions, assuming a power-law source function with $\chi_S<-1$, which is given by Equation \ref{eq:n_solution_power_law_chis_neq_minus_1}. Removing the common normalization factor $n_0$, the number-density model has four free parameters: $\chi_\tau,\,\chi_S,\,P_\mathrm{cut}$ and $P_\mathrm{mig}$. We numerically integrate Equation \ref{eq:n_solution_power_law_chis_neq_minus_1} in the corresponding log-period bin using \texttt{numpy.trapezoid} to obtain the model occurrence rate corresponding
to the binned occurrence rate obtained from the data. The total rates from both the data and the model are normalized to 1 by summing over every bin in the data range.

To characterize the constraints and degeneracies of the three classes of solutions mentioned above, we sample their parameter posteriors using \texttt{emcee} \citep{emcee,emcee2}. Because the three solutions occupy distinct regions of parameter space, we sample them separately by artificially enforcing different bounds for the parameters. This is equivalent to adopting uniform priors within the specified parameter ranges and zero probability outside them, and it helps to constrain the individual classes of solutions instead of a full parameter-space search that depends heavily on the prior.

The slopes seen in the data indicate $d\log\tilde{n}/d\log P\simeq 3$ for periods
below the turnover at $\simeq 3$ days, and $d\log\tilde{n}/d\log P\simeq -1$ for longer
periods. Hence, for solution 1, we restrict $2.5\leq\chi_{\tau}\leq5$, encompassing
the expected value $\chi_\tau\simeq 3$. We also restrict $-10\leq\chi_{\rm S}\leq-1.01$ since we consider only the regime $\chi_{\rm S}<-1$. $P_\mathrm{cut}$ and $P_\mathrm{mig}$ are restricted to lie between $0.4\;\mathrm{days}$ and $10\;\mathrm{days}$, the range of periods probed by the data.

For solution 2, the apparent power-law index of $\simeq 3$ for periods below the turnover suggests $\chi_\tau<3$. One can see this from the numerical solutions in Figure \ref{fig:power-law-s}, that the dark brown curves all appear steeper than the blue curves ($\propto P^{\chi_\tau}$) when $P<P_\mathrm{cut}$ (see also the formal derivation in Appendix \ref{appendix:truncated_power_law}, after Equation \ref{eq:chi_I_general_when_chi_S_neq_-1}). Therefore, we restrict $0.1\leq\chi_{\tau}\leq2.5$ for the second solution, and keep the bounds for other parameters the same as in the first solution.

Under the third solution,
the power-law index of 3 observed at short periods
implies $\chi_\tau+\chi_S+1\simeq 3$.
Since $\chi_S+1<0$, this solution
favors $\chi_\tau>3$. The solution can also be characterized as requiring $P_\mathrm{cut}$ to be shorter than the observed turnover period. We therefore restrict $2.5\leq\chi_{\tau}\leq8$ and change the $P_\mathrm{cut}$ range to $0.4\;\mathrm{days}\leq P_{\rm cut}\leq1\;\mathrm{day}$. The other parameter ranges are the same as in solution 1.

For each solution, we first perform a multi-start, bounded local optimization from a grid of initial parameter values. The converged solutions with $\chi^2$ values within 5\% of the lowest value are used to determine a common starting point, about which we randomly initialize 32 walkers with small random perturbations. This preliminary optimization is used only to initialize the ensemble. We then evolve each walker for 1 million steps and discard the first $5\times10^4$ steps, retaining every 1$,$000th sample. The largest estimated autocorrelation time is approximately 350 steps, such that the chains span many autocorrelation times after burn-in, indicating adequate convergence.

We then identify the highest-likelihood MCMC sample and use it as the initial point for a final bounded local optimization, thereby obtaining the maximum-likelihood parameters for the three solutions. We show the solutions with these parameters in Figure \ref{fig:fit}, where the shaded regions show the 95\% credible intervals of the model, calculated from the 2.5th and 97.5th percentiles of the model predictions generated from the MCMC samples. The three solutions correspond to i) $\chi_\tau=3$ with $P_\mathrm{mig}>P_\mathrm{cut}=3.7\,\mathrm{days}$, ii) $\chi_\tau=1.7$ with $P_\mathrm{mig}\approx P_\mathrm{cut}=3.8\,\mathrm{days}$, and iii) $\chi_\tau=5.6$ with $P_\mathrm{cut}<P_\mathrm{mig}=4.4\,\mathrm{days}$, respectively, consistent with the three qualitative regimes identified previously.

\begin{figure*}
    \centering
    \includegraphics[width=0.495\linewidth]{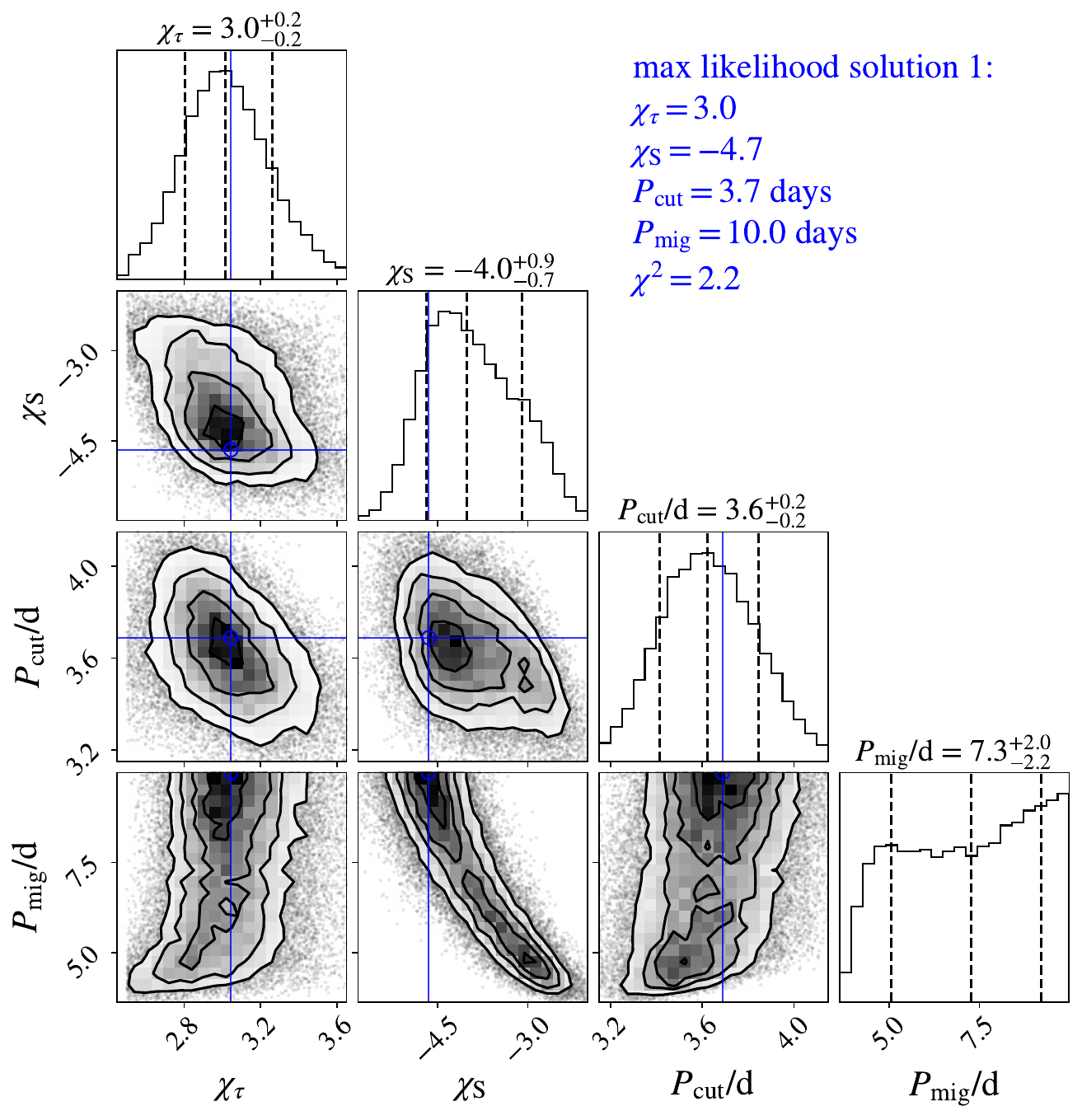}
    \includegraphics[width=0.495\linewidth]{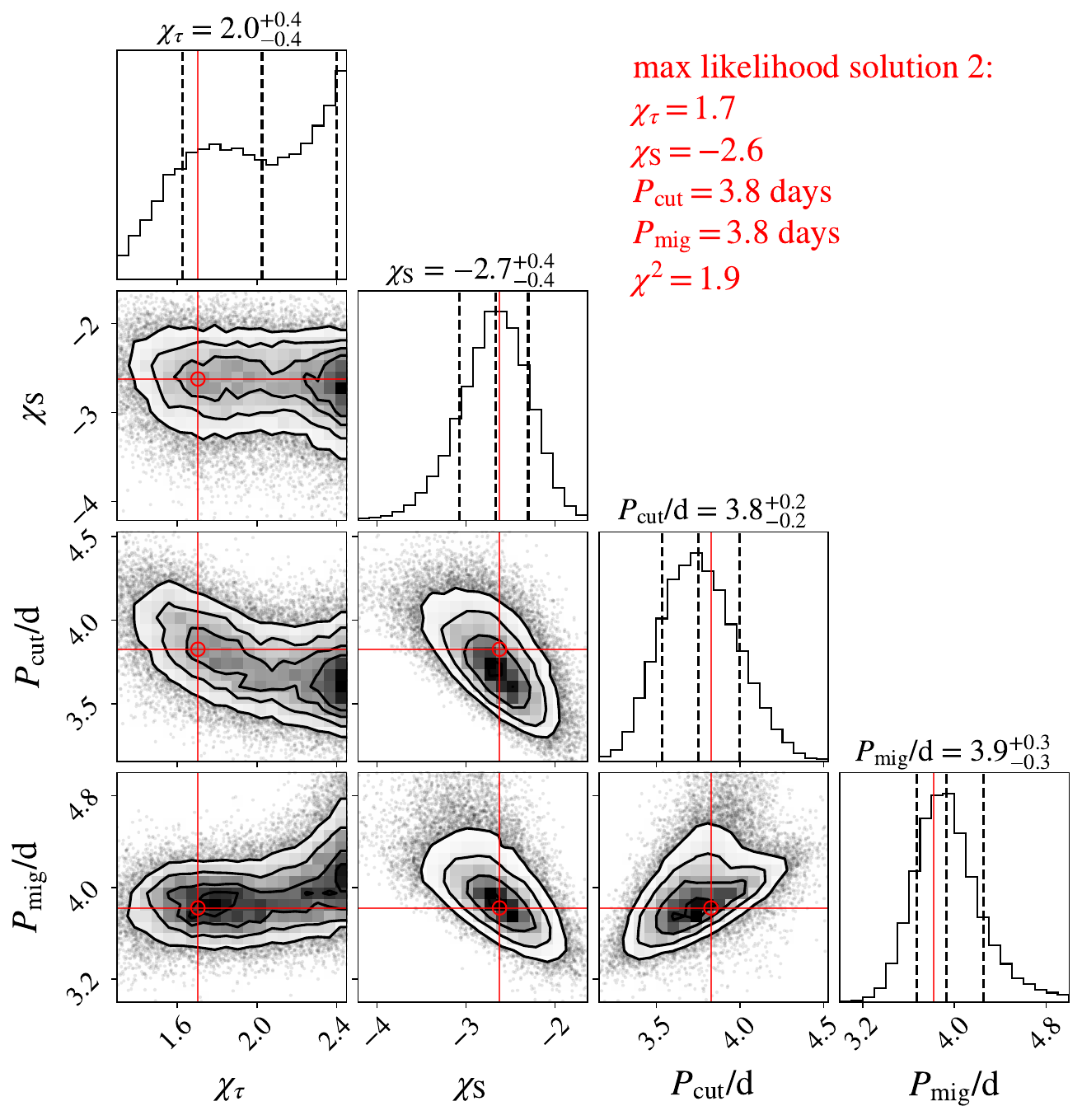}

    \includegraphics[width=0.495\linewidth]{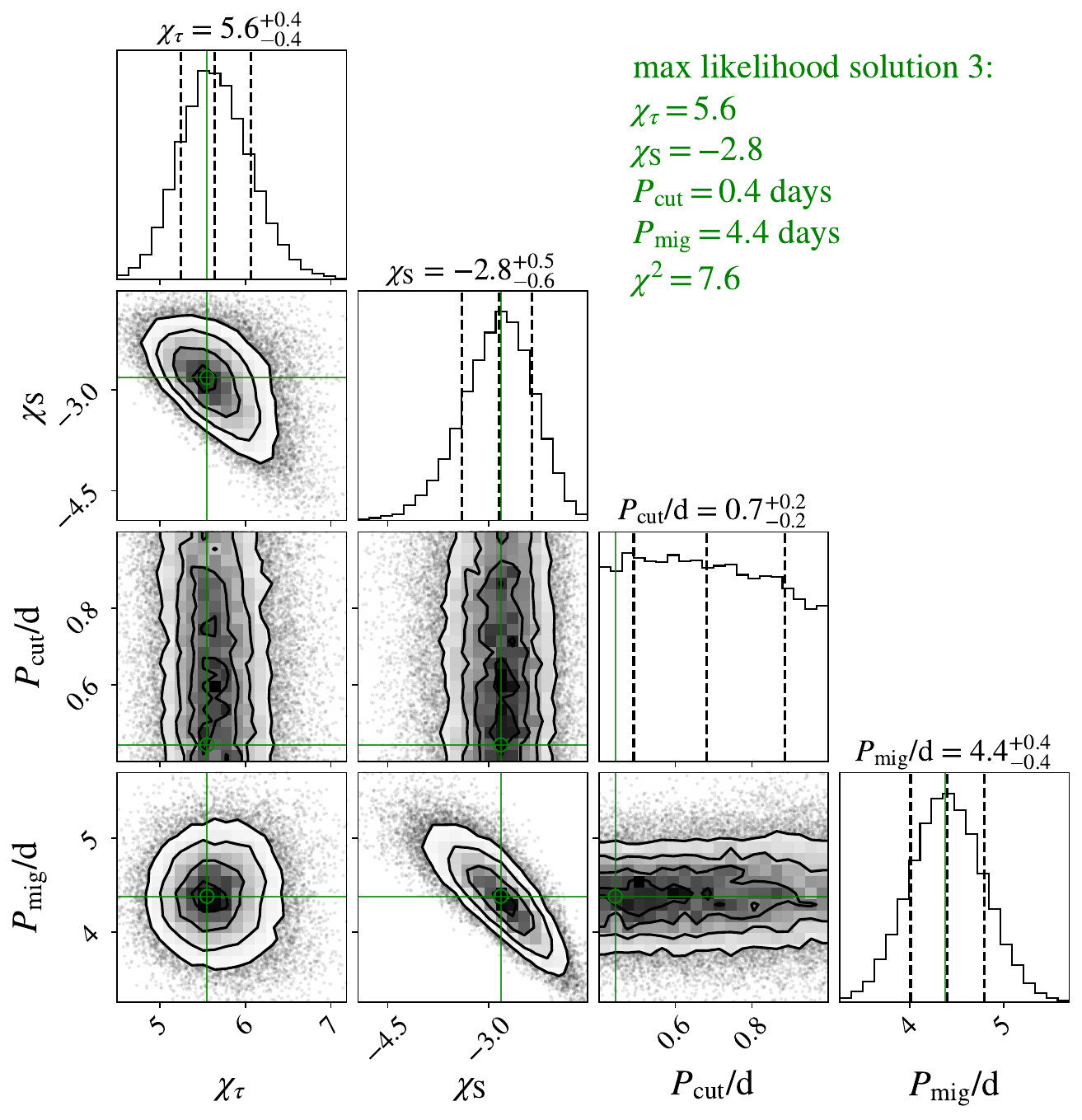}
    \caption{
        Corner plots showing the MCMC posteriors. The quoted parameter ranges correspond to the 16th, 50th, and 84th percentiles. The maximum-likelihood parameters are identified with colored open circles, horizontal lines, and vertical lines.
        All three solutions yield acceptable $\chi^2$ values.
        Solution 1 gives only a lower limit on $P_{\rm mig}$; the posterior extends to the imposed cutoff of $10\;\mathrm{days}$. In solution 2, $\chi_\tau$ is not well constrained, and the constraint $P_{\rm cut} \approx P_{\rm mig}$ is visible as a correlation in the 2-d posterior. In solution 3, $P_{\rm cut}$ is not well constrained; the posterior extends to the imposed cutoff of $0.4\;\mathrm{days}$.
    }
    \label{fig:corners}
\end{figure*}

We show the resulting MCMC posterior distributions in Figure \ref{fig:corners}, where the quoted parameter ranges correspond to the 16th, 50th, and 84th percentiles. The optimized parameters that correspond to the solutions in Figure \ref{fig:fit} are also overlaid on the posterior distributions as colored open circles and horizontal and vertical lines.

For solution 1 (blue, first panel), the posterior of $P_{\rm mig}$ extends toward the imposed upper bound and is therefore only weakly constrained on the upper side. This is expected in our qualitative analysis because a third power law at $P>P_\mathrm{mig}$ is not identified in the data.
Although $\chi_\tau$ is well
constrained, a significant degeneracy exists between $\chi_S$ and $P_{\rm mig}$. This degeneracy arises because the power-law behavior at $P\gtrsim3.5\;\mathrm{days}$ depends jointly on $\chi_{\tau}$ and $\chi_S$, as well as on the locations of $P_{\rm cut}$ and $P_{\rm mig}$, as illustrated by the numerical examples in the left panel of Figure \ref{fig:power-law-s}. 

For solution 2 (red, second panel), the property $P_\mathrm{mig}\approx P_\mathrm{cut}$ is visible as a correlation in the posterior, as expected. However, $\chi_\tau$ is not well constrained, and its posterior appears to be flat. This is because, in this solution, the period dependence of number density is not a power-law function and has a complicated dependence on $\chi_\tau$ (see discussions in Appendix \ref{appendix:truncated_power_law}). The $\chi_\tau$ posterior also extends to the enforced upper bounds at $2.5$, which confirms that this solution extends to the first solution when $P_\mathrm{mig}$ varies continuously.

For solution 3 (green, third panel), $\chi_\tau$ is well constrained.
In contrast, the posterior of $P_{\rm cut}$ extends toward the imposed lower bound and is only weakly constrained on the lower side. This is again expected in our qualitative analysis because a third power law at $P<P_\mathrm{cut}$ is not identified in the data. A similar degeneracy is also present between $\chi_S$ and $P_{\rm mig}$ for this solution.

All three solutions have acceptable $\chi^2$ values given $n_\mathrm{DOF}=5$. Thus,
we conclude that all three
solutions are consistent with the data. Solution 2 requires $P_\mathrm{mig}\approx P_\mathrm{cut}$, a coincidence
for which we do not have a physical explanation,
making this solution appear finely-tuned.

\section{Discussion}
\label{sec:discussion}

\subsection{Tidal Migration Theories}
\label{sec:migration_theories}

Below, we discuss several
previously studied
tidal migration theories for hot Jupiters and compare their predictions with the values of $\chi_\tau$ and $P_\mathrm{mig}$ inferred from the data. Sometimes, the tidal migration is parameterized by the effective tidal quality factor $Q'$, defined as \citep{goldreich1966}:
\begin{equation}    
\label{eq:tidal_Q}
Q'\equiv\frac{27\pi}{2} \frac{ M_\mathrm{p}}{M_\mathrm{\star}}\left(\frac{R_\mathrm{\star}}{a}\right)^5\left(\frac{1}{P}\right)\left(\frac{P}{|\dot{P}|}\right)\,,
\end{equation}
where $M_\mathrm{\star}$ and $M_\mathrm{p}$ are the masses of the host star and the planet, and $R_\mathrm{\star}$ and $a$ are the radius of the host star and the semi-major axis of the orbit. This expression neglects the rotation of the host star. With our parameterization of tidal migration in Equation \ref{eq:tau_mig}, this leads to
\begin{equation}
\label{eq:tidal_Q_dependence_on_P}
    Q'\propto  P^{\chi_\tau-13/3}\,,
\end{equation}
where the $-13/3$ factor comes from the $a^{-5}P^{-1}$ dependence in the definition of $Q'$.

\cite{Ma2021} studied a scenario in which a planet's forcing frequency is synchronized
with an evolving eigenfrequency of an
internal gravity-wave mode of
the host star.
This scenario, known as ``resonance locking,'' predicts that the tidal migration timescale is independent of the orbital period, i.e., $\chi_\tau=0$. This is because the star's natural oscillation frequencies evolve independently of the planet. As we discussed in Section \ref{sec:res_lock}, the observed number density generally depends on the source function when $\chi_\tau=0$. Our best-fit solutions are not consistent with $\chi_\tau=0$, thereby ruling
out resonance locking under the
assumption that the source function is a truncated
power law. Note, though, that
resonance locking cannot be excluded
for arbitrary source functions (see Section~\ref{subsec:millholland_et_al}).

The classical theory of equilibrium tides has been widely applied to the study of orbital decay of hot Jupiters. In this picture, the planet raises a tidal bulge that dissipates energy through turbulent viscosity in the convective stellar envelope \citep{goldreich1966}. When the dissipation is efficient, it predicts a tidal quality factor independent of period (therefore also known as the ``constant $Q$ theory'')
in which $\chi_\tau=13/3\approx4.33$,
based on Equation \ref{eq:tidal_Q_dependence_on_P}. This value is higher than the values we found from our solutions 1 and 2 and lower than that from solution 3. For a Jupiter-mass planet orbiting a solar-mass star, this theory gives $P_\mathrm{mig}\approx 4\,\mathrm{days}$ assuming $Q'=10^6$ and $t=8\,\mathrm{Gyr}$.

Nevertheless, it has been argued that when the tidal forcing period is short (hours to days), the dissipation of equilibrium tides could be suppressed (see, e.g., \citealt{Zahn1966,Goldreich1977,Goldreich1977b,Zahn1989,Goodman1997,Ogilvie2012,Duguid2020,Duguid2020b,Vidal2020}). In this regime, the excitation of tidal bulges occurs on a timescale shorter than the turnover timescale of turbulent eddies\footnote{The turnover timescale is $\sim20\,\mathrm{days}$ in the middle of the convective envelope for sun-like stars, and it increases with the depth of the convective zone.}, causing the turbulent dissipation efficiency to be reduced. This scenario, which we refer to as ``fast tides suppression'' below, would imply weaker dissipation for shorter period orbits compared to the constant $Q$ model.  

Specifically, \cite{Duguid2020b} find that the effective turbulent viscosity scales as $\nu_\mathrm{eff}\propto(P_\mathrm{f}/\tau_\mathrm{conv})^{2}$ for the periods corresponding to hot Jupiter orbital decay, where $P_\mathrm{f}$ is the tidal forcing period. With this scaling, the tidal quality factor scales as $Q'\propto P^{-1}$, which leads to $\chi_\tau=10/3\approx 3.3$. This is probably consistent with solution 1 depicted in Figure \ref{fig:fit}, in which $\chi_\tau=3.0\pm 0.2$. However, the viscous strength they found from numerical simulations would produce an extremely long orbital decay timescale, leading to $P_\mathrm{mig}\approx 0.4\,\mathrm{days}$ assuming the same parameters as before, far below the inferred lower-limit in our solution 1. Nevertheless, because the estimate of viscous strength is subject to large uncertainty, the estimate for $P_\mathrm{mig}$ is probably not as robustly constrained as $\chi_\tau$. Hence, although the ``fast tides suppression'' scenario is compatible with the solution 1
in its value of $\chi_\tau$, future work should investigate whether it can also
produce a consistent value of $P_\mathrm{mig}$.

\cite{Terquem2021} has proposed a new picture that challenges the idea of fast tides suppression. \cite{Terquem2021} argues that when planets excite stellar tides on timescales shorter than the convective turnover time, the dominant terms that characterize the interaction between tidal flows and convection are different from what has been previously assumed (see, e.g., related discussions and numerical simulations by \citealt{Barker2021,Terquem2021comment,Terquem2026,Zhou2026}). In this picture, which we refer to as ``fast tides enhancement'', dissipation at short periods would be enhanced rather than suppressed. The numerical results of
\cite{Terquem2021} suggest a
migration law with $\chi_\tau=5.7$ and $P_\mathrm{mig}\approx 5\,\mathrm{days}$, assuming the same parameters as before. If true, this scenario is compatible with our model under solution 3.

Another tidal dissipation mechanism is the weakly nonlinear dissipation of internal gravity waves of the host star. These waves, when tidally excited by a planet's orbit, can reach significant amplitudes and dissipate their energy
through nonlinear wave coupling. \cite{Essick2016} showed that for hot Jupiters on hour to day orbits around sun-like stars, nonlinear mode coupling is
expected to yield $\chi_\tau\approx6-7$ and $P_\mathrm{mig}\approx 3\,\mathrm{days}$, assuming the same parameters as before. The inferred $\chi_\tau$ is somewhat higher than the range of
values allowed under solution 3, but it might be deemed compatible given the simplifications inherent in our model.

When tidally-excited gravity waves reach very high amplitudes, they trigger nonlinear wave breaking and efficiently dissipate their energy through turbulent viscous flows. \cite{Barker2010} showed that if nonlinear wave breaking occurs, $Q'\propto P^{8/3}$, implying $\chi_\tau=7$ and $P_\mathrm{mig}\approx 3\,\mathrm{days}$, assuming the same parameters as before. While $P_\mathrm{mig}$ seems compatible, the implied $\chi_\tau$ is higher than the values we inferred in any of our solutions. Therefore, this theory would predict
even fewer planets at short periods than are observed.

In summary, our solution 1 with $\chi_\tau=3$ seems to be consistent with the ``fast tides suppression'' theory, where planets migrate through frequency-dependent suppressed dissipation of equilibrium tides that they raise on their hosts, though the induced $P_\mathrm{mig}$ is not entirely consistent with the solution. We are not aware of any previously proposed tidal theory that is consistent with our solution 2, with $\chi_\tau=1.7$. Our solution 3, with $\chi_\tau=5.6$, is consistent with theories that produce rapid migration at short periods, such as ``fast tides enhancement'' and weakly nonlinear dissipation of waves. 

\subsection{Source Function Theories}

\begin{figure}
    \centering
    \includegraphics[width=\columnwidth]{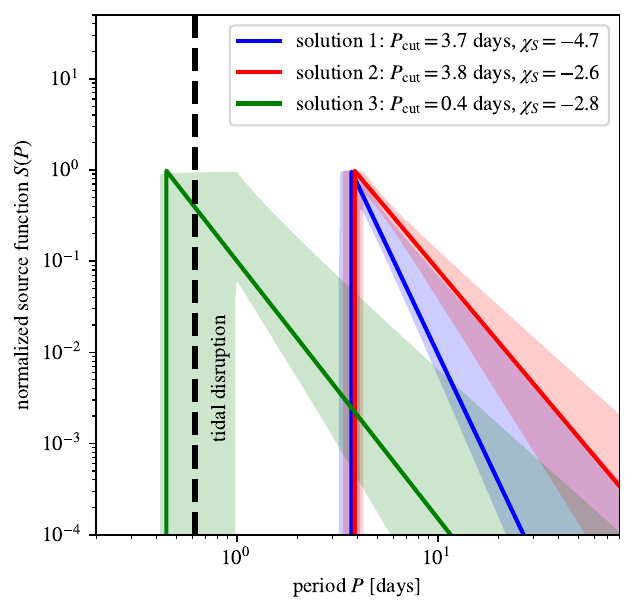}
    \caption{Shown are the source functions (Equation \ref{eq:powerlaw_source}) implied by our comparison of theory to data, normalized to unity at the lower truncation period $P=P_\mathrm{cut}$. The colored lines are based on the maximum-likelihood parameters. The shaded regions show the 95\% credible intervals. Solutions 1 and 2 imply $P_\mathrm{cut}\sim 4\,\mathrm{days}$, consistent with theoretical predictions of high-eccentricity migration. In solution 3, the source function extends all the way down
    to the tidal disruption period (black dashed line), which seems unlikely based on current theories for hot Jupiter formation.
    }
    \label{fig:source}
\end{figure}

Each of the 3 solutions we find involves a different source function, as shown in Figure \ref{fig:source}. The colored lines show the source functions (Equation \ref{eq:powerlaw_source}) calculated from the maximum-likelihood parameters for each solution, and the colored regions show the 95\% credible intervals drawn from their corresponding posteriors. All source functions are normalized to $S=1$ at $P=P_\mathrm{cut}$.

Solutions 1 and 2 both have lower truncation periods of $P_\mathrm{cut}\approx 3.7\,\mathrm{days}$, and a power-law
fall-off at longer periods. The truncation periods are well constrained in these two solutions, as they correspond to the turnover of number density in the data, as one can see in Figure \ref{fig:fit}. These source functions imply that hot Jupiter
circularization typically results
in periods near $\simeq 3.7\,\mathrm{days}$.

This finding is consistent with theoretical models of hot Jupiter circularization through high-eccentricity migration. Specifically, \cite{Wu2018} argued that hot Jupiters stall at a periastron distance $D_\mathrm{p}\sim 4r_\mathrm{t}$ before they are quickly circularized by f-mode diffusion, where $r_\mathrm{t}=(M_\mathrm{\star}/M_\mathrm{p})^{1/3}\,R_\mathrm{p}$ is the tidal radius and $R_\mathrm{p}$ is the planet radius. In this theory, the circularized hot Jupiters pile up around a semi-major axis $a=2D_\mathrm{p}$, assuming that the orbital angular momentum is conserved during circularization. If we assume $M_\mathrm{\star}=M_\odot$, $M_\mathrm{p}=M_\mathrm{Jupiter}$, and $R_\mathrm{p}=1.2\,R_\mathrm{Jupiter}$ (to account for the thermally induced
``inflation'' of hot Jupiters), the pile-up would occur at $a\simeq 0.05\,\mathrm{AU}$ or $P\simeq 4\,\mathrm{days}$, which agrees well with the lower truncation period of the source function associated with solutions 1 and 2.

Solution 3, on the other hand, requires
a source function
that extends down to periods
shorter than one day. While it has been found that planets on highly eccentric orbits can, in principle, reach a periastron distance $D_\mathrm{p}$ close to the Roche limit \citep{Rappaport2013}, this seems unlikely to apply to the majority of hot Jupiters and cannot easily explain the peak at sub-day periods implicit in this solution. 
In this sense, solution 3
is disfavored by requiring a source
function that seems physically
implausible.

\subsection{Constraints from Individual Systems}
\label{sec:individual_migration_rate}

Solutions 1--3 are all statistically
consistent with the present-day
period distribution of hot Jupiters,
but they make different predictions
about the actual migration rate $|\dot{P}|$ 
at any given period.
Thus, the detection or upper limits
on period changes for individual hot Jupiters may provide additional constraints on our solutions. Specifically, for given choices of parameters $\chi_\tau$, $P_\mathrm{mig}$, and $t$, the rate of period
decay is given by Equation \ref{eq:tau_mig_P_mig}:
\begin{equation}
\label{eq:dotP}
    |\dot{P}|=\frac{P}{\chi_\tau t}\left(\frac{P}{P_\mathrm{mig}}\right)^{-\chi_\tau}\,.
\end{equation}

\begin{figure}
    \centering
    \includegraphics[width=\columnwidth]{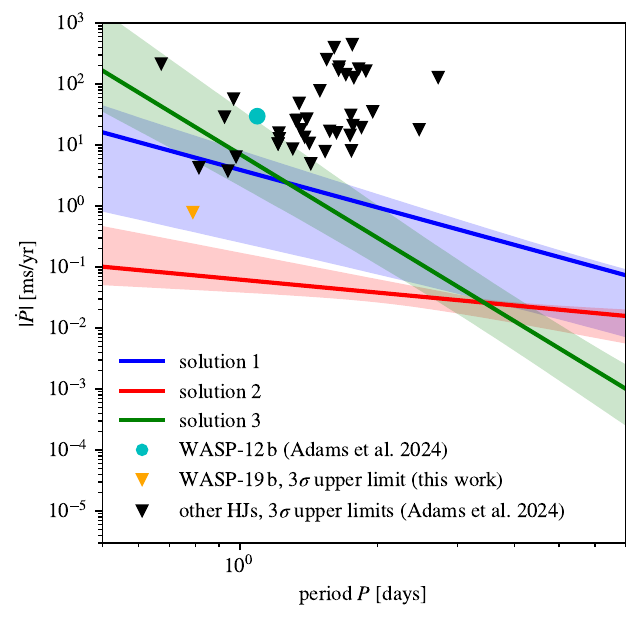}
    \caption{Tidal migration rates inferred from our model (colored lines) and as constrained for individual systems (cyan dot for WASP-12\,b and triangles for $3\sigma$ upper limits). The colored lines
    are based on the maximum-likelihood  parameters of the 3 solutions and the shaded regions show the 95\% credible intervals. The rates for individual systems are from Table 7 of \protect\cite{Adams2024}, with our updated value obtained for WASP-19\,b (orange triangle). Only solution 3 is marginally consistent with WASP-12\,b, though there are reasons why this
    system may be exceptional (see Section~\ref{sec:individual_migration_rate}). The non-detection of orbital decay for
    WASP-19\,b favors solution 2. The other upper limits are generally not strong enough to constrain the models, though a few systems disfavor solution 3.
    }
    \label{fig:decay_compare}
\end{figure}

Figure \ref{fig:decay_compare} shows the inferred period decay rate from our three solutions with colored lines, assuming $t=8\,\mathrm{Gyr}$
and using our maximum-likelihood parameters from each solution. 
Note that the choice of $t$ is somewhat
arbitrary, as it cannot be inferred
from the data due to the degeneracy with
the overall normalization of the number
density.
The shaded regions in the figure show the 95\% credible intervals drawn from the MCMC posteriors for these solutions, assuming a flat prior for $t$ between 6 and 10 Gyr. Despite the large spread in predictions for $|\dot{P}|$, our three solutions predict order-of-magnitude differences in migration rates for hot Jupiters with periods shorter than a few
days.

For comparison, Figure \ref{fig:decay_compare}  also shows constraints obtained for individual hot Jupiters, based on the compilation of results for
43 systems given 
in Table 7 of \cite{Adams2024}\footnote{We
excluded TrES-1\,b, whose orbital decay has not been firmly established; see, e.g., \cite{Sodickson2025}.}. Among them, WASP-12\,b is the only known hot Jupiter with compelling evidence of orbital decay \citep{Patra2017,Maciejewski2016,Yee2020}. The measured
rate of  $|\dot{P}|=29.8\pm1.6\,\mathrm{ms/yr}$ is shown as a cyan dot in Figure \ref{fig:decay_compare}. WASP-19\,b, on the other hand, has an even shorter period
(0.78~d) and has been monitored nearly
as long as WASP-12\,b, yet has not
been seen to decay.
To facilitate comparison between the upper
limits on its decay and our model,
we re-determined the orbital ephemerides of WASP-19\,b (see Appendix \ref{app:wasp19b}).
The result is a $3\sigma$ upper limit of $|\dot{P}|<0.80\,\mathrm{ms/yr}$
depicted by the orange triangle
in Figure \ref{fig:decay_compare}. 
The black triangles are $3\sigma$ upper limits for other systems drawn from \cite{Adams2024}, none of which
is as constraining as WASP-19\,b.

The fact that a decay rate as fast as
WASP-12\,b's can be ruled out
for WASP-19\,b is evidence that these
systems differ in a way that
is not reflected in our model.
Furthermore, the migration rate measured for WASP-12\,b is high enough to be outside most of the posteriors from all 3 solutions, being only marginally consistent with solution 3. This
could be interpreted as evidence favoring
solution 3, or that WASP-12\,b is
an exceptional case.
A widely discussed possibility is that the host star of WASP-12\,b has evolved off
the main sequence
(\citealt{Weinberg2017,Bailey2019}, though see \citealt{Efroimsky2022,Leonardi2024,Golonka2026} for counter-arguments), and tidal dissipation
in evolved stars may be very different
from main-sequence stars. Other explanations include physical processes not related to tidal decay, such as the possibility
that WASP-12\,b is losing orbital
angular momentum to an escaping wind,
although the details of this process
remain uncertain \citep{Valsecchi2015,Jia2017,Jackson2017,Yee2020}.

On the other hand, there appears
to be nothing exceptional about WASP-19\,b, given the $0.97\,M_\odot$ host star mass and $1.15\,M_\mathrm{Jupiter}$ planetary mass measured \citep{Cortes-Zuleta2020}. It might
therefore be a representative case of hot Jupiter tidal migration. The empirical
upper limit on $|\dot{P}|$ favors solution 2, as most of the posterior density in solutions 1 and 3 lies above this limit. The upper limits for a few other systems 
also disfavor solution 3, though not
as strongly.
For most other hot Jupiters, the upper limits on tidal migration achieved to date do not provide meaningful constraints on our solutions.

One caveat to the comparison in Figure \ref{fig:decay_compare} is that our parameterization of tidal migration (Equation \ref{eq:tau_mig}) should be understood only as an overall long-term average of the actual migration rate. The instantaneous migration rates of hot Jupiters, which are the ones measured, may not be a smooth, monotonic function of period, as we assumed. This is because when hot Jupiters tidally excite stellar oscillation modes, they may cross or leave resonances that cause
much faster or slower migration than the long-term average \citep{Bryan2024,Wu2024}. If this is the case, the individual migration rates measured may not constrain the time-averaged solutions well because it is hard to know whether the planets are on resonances or not. This may explain the discrepancies between WASP-19\,b measurements and our solutions 1 and 3, as this hot Jupiter could be off resonance and is currently migrating much more slowly than its long-term average. It may also provide another explanation for the high migration rate obtained for WASP-12\,b, other than post-main-sequence tides, as it could well be migrating around a sun-like host but happens to be on-resonance.

In summary, solution 1 is consistent with most of the upper limits of $|\dot{P}|$ obtained for known hot Jupiters, but not with WASP-19\,b and the fast-migrating WASP-12\,b. Solution 2, which predicts a slower migration rate at short periods, is  consistent with the upper limit for WASP-19\,b but not with the WASP-12\,b measurement. Solution 3 is disfavored by the upper
limits on a few short-period hot Jupiters but is marginally consistent with the migration rate of WASP-12\,b. Post-main-sequence tides and resonant migration, which are not included in our model assumptions, may explain the discrepancy between the measurements and our model predictions. 

\subsection{Implied Hot Jupiter Destruction Rate}
\label{sec:destruction}

Tidally migrating hot Jupiters can ultimately be disrupted or engulfed by their host stars, which may explain the observational evidence that those stars tend to be younger (\citealt{Hamer2019,Mustill2022,Miyazaki2023,Chen2023,Banerjee2024}; see also \citealt{Hu2026} for a counterargument). The destruction of hot Jupiters may cause transient events that can be observed, and leave traces in the stellar photosphere with unique trends of chemical abundances \citep{Behmard2023}. \cite{De2023} and \cite{Lau2025} recently presented observations of the subluminous red nova ZTF SLRN-2020 and interpreted it as a planet engulfed by its host star through tidal orbital decay on the main sequence. Our solutions, along with the normalization provided by the data, can give an estimate of the rate of such events.

Assuming that the migrating hot Jupiter is eventually tidally disrupted at the Roche radius $R_\mathrm{Roche}=2.44\,R_\mathrm{t}$, the rate of such an event $\Gamma$ is given by the flux of planets that is crossing through the period $P_\mathrm{d}$ that corresponds to $R_\mathrm{Roche}$:
\begin{equation}
    \Gamma=n|\dot{P}|\;\mathrm{evaluated\;at\;}P_\mathrm{d}\,.
\end{equation}
Here, $n$ is normalized by the observed hot Jupiter occurrence rate $n_\mathrm{ref}$ at a reference period $P_\mathrm{ref}$. The data we obtained from Yee et al. (in prep) give a rate of $\approx 6\times10^{-5}$ per sun-like star around $P_\mathrm{ref}=0.875\,\mathrm{days}$ in a one-day period bin. Assuming there are $\sim5\times10^{10}$ sun-like stars in the Galaxy, this gives a Galactic hot Jupiter period-space number density of $n_\mathrm{ref}\sim3\times10^6\;\mathrm{d}^{-1}$ at $P\simeq P_\mathrm{ref}$.

We obtain the hot Jupiter Galactic disruption rates based on the maximum-likelihood parameters from our three solutions. We assume a Jupiter-like planet orbiting a sun-like star, which gives $P_\mathrm{d}=0.6\,\mathrm{days}$ corresponding to $R_\mathrm{Roche}=0.014\,\mathrm{AU}$, when taking $M_\mathrm{\star}=M_\odot$, $M_\mathrm{p}=M_\mathrm{Jupiter}$, and $R_\mathrm{p}=1.2\,R_\mathrm{Jupiter}$ to account for the slight inflation of hot Jupiters. With the above parameters and assuming $t=8\,\mathrm{Gyr}$, solution 1 gives $\Gamma_{1}\sim0.2\,\mathrm{yr}^{-1}$. Solution 2 gives a much lower value, with $\Gamma_{2}\sim10^{-3}\,\mathrm{yr}^{-1}$. Solution 3, on the other hand, gives $\Gamma_{3}\sim1\,\mathrm{yr}^{-1}$. As some of the parameters are not well constrained in the above solutions, we expect an uncertainty of a factor $\sim\mathcal{O}(10)$ in the above values.

\cite{De2023} estimated the Galactic rate of subluminous red novae potentially associated with planetary destruction by simulating mock observations that were selected in the same way as ZTF-SLRN-2020 is identified. They obtained $\Gamma=4.3^{+7.5}_{-3.1}\,\mathrm{yr}^{-1}$ with 68\% confidence intervals. If all these events are caused by hot Jupiter destruction, this rate is roughly consistent with the model predictions from our solutions 1 and 3, given the uncertainties. Solution 2, on the other hand, likely predicts that tidal
destruction is unlikely to be observed
on human timescales.  

\subsection{Comparison to Millholland et al. 2025}
\label{subsec:millholland_et_al}
Our work was stimulated by the approach taken by
\citealt{Millholland2025} (M25) to inferring the tidal migration law for hot Jupiters,
and their inference of a strongly period-dependent $Q'$
consistent with tidal resonance locking.
Ultimately, though,
we reached a different conclusion about what
can be inferred from the present-day population. While M25 found a broad range of $0\lesssim\chi_\tau\lesssim2.33$ ($-4.33\lesssim\alpha\lesssim-2$ in their parameterization) that is consistent with the data, our analysis yields three solutions with different preferred values of $\chi_\tau$. Specifically, the case of resonance locking ($\chi_\tau\approx0$) is not supported by the data according to our analysis.

The disagreement exists on two levels.
First, if we adopt the truncated power-law source function with
$\chi_S < -1$, Section~\ref{subsec:data_fitting} shows that
the observed period distribution favors $\chi_\tau > 0$
and excludes the resonance-locking case of $\chi_\tau\approx 0$.
Second, and more fundamentally,
if the form of the source function is left unconstrained,
the observed period distribution cannot uniquely test the
resonance-locking hypothesis. This is because under the resonance-locking
hypothesis, all planets experience a steady decrease in $\ln P$
and the observed number density always retains a direct
dependence on the source function (see Section~\ref{sec:res_lock}).
Thus, one can always obtain agreement with the data by choosing a suitable source function.
Without independent knowledge of the source function,
the amount of tidal migration via resonance locking cannot be inferred uniquely.

What is responsible for these different conclusions?
Our analysis makes use of the full time-dependent solution in period space, whereas M25's inference
was based primarily on the approximately
steady-state portion of the decay-time distribution.
We were also able to benefit from an updated hot-Jupiter period distribution.
However, the key difference is probably
the parameterization of the number
density function. Instead of the number of planets per period interval $dP$,
M25 used the number of planets per decay-time interval $d\tau_d$,
where $\tau_d$ is defined as the time to migrate to $P=0$ under the
assumed tidal migration law. This parameterization
has the advantage of casting the continuity equation
as a simple constant-velocity advection equation
(see Eq.~8 of M25 and Eq.~\ref{eq:continuity_tau_appendix}
in Appendix \ref{sec:number_density_expression}).
However, the transformation from
$P$ to $\tau_d$ becomes
increasingly insensitive to $\chi_\tau$ as
$\chi_\tau \rightarrow 0$, making it
difficult to interpret the constraints as $\chi_\tau$ approaches the resonance-locking limit.

Specifically, M25 modeled the $\tau_d$-space number density as a power
law of slope $p$ (M25, Eq. 17):
\begin{equation}
    p=\frac{1-(\tau_d/\tau_{d,p})^\gamma}{1-\frac{1}{\gamma+1}(\tau_d/\tau_{d,p})^\gamma}\,.
\end{equation}
Here, $\tau_{d,p}>\tau_d$ is a normalization of $\tau_d$ defined from the steady state solution. The index $\gamma=2/(13+3\alpha)$ is related to M25's parameterization of the tidal migration rate (M25, Eq. 21):
\begin{equation}
    Q'\propto P^\alpha\,,
\end{equation}
where $Q'(P)\propto \tau_\mathrm{mig}P^{-13/3}$ is the reduced tidal quality factor. It is then straightforward to obtain the relations between $\gamma$, $\alpha$, and our parameterization $\chi_\tau$:
\begin{equation}
    \alpha = \chi_\tau-\frac{13}{3};\;\gamma=\frac{2}{3\chi_\tau}\,.
\end{equation}
This leads to the relationship between M25's fitting parameter $p$ and our parameter $\chi_\tau$:
\begin{equation}
    p=\frac{1-(\tau_d/\tau_{d,p})^{\frac{2}{3\chi_\tau}}}{1-\frac{3\chi_\tau}{3\chi_\tau+2}(\tau_d/\tau_{d,p})^{\frac{2}{3\chi_\tau}}}\,.
\end{equation}
M25 fitted the observed number density in $\tau_d$ space to constrain $p$, which in turn provides constraints on $\alpha$ (and hence $\chi_\tau$). However, the parameter $p$ becomes insensitive to $\chi_\tau$ as $\chi_\tau\rightarrow0$ (the resonance-locking case), as
can be seen by taking the derivative of $p$ with respect to $\chi_\tau$:
\begin{equation}
    \frac{d p}{d\chi_\tau} =
\frac{
2\epsilon^{\frac{2}{3\chi_\tau}}
\left[
9\chi_\tau^2
\left(1-\epsilon^{\frac{2}{3\chi_\tau}}\right)
+
2(3\chi_\tau+2)\ln\epsilon
\right]
}{
3\chi_\tau^2(3\chi_\tau+2)^2
\left[
1-\dfrac{3\chi_\tau}{3\chi_\tau+2}
\epsilon^{\frac{2}{3\chi_\tau}}
\right]^2
}\,,
\end{equation}
where $\epsilon\equiv \tau_d/\tau_{d,p}<1$.
As $\chi_\tau\rightarrow0^+$, the derivative vanishes.
In this
regime, small errors in the estimate of $p$ are magnified into
large errors in $\chi_\tau$. Consequently, even a well-constrained value of $p$ may translate into a weak constraint on $\chi_\tau$
near the resonance-locking limit.

\subsection{Assumptions and Future Improvements}
\label{sec:caveats}

In this work, we make a number of assumptions for simplicity. Nevertheless, the analytical formalism we developed can be generalized when these assumptions are relaxed, as we discuss below.

The most fundamental assumption of our
model is that every hot Jupiter migrates in the same way: we assumed the migration rate is a universal function of $P$ (Equation \ref{eq:tau_mig}). This is inspired by the fact that the observed sample of hot Jupiters is a fairly homogeneous population with similar planetary and stellar properties. However, tidal migration theories sometimes predict that
varying the system parameters even slightly
can lead to drastically different dissipation
rates. For example, the possibility
of nonlinear dissipation of
internal gravity waves 
depends sensitively on whether the host star is hotter or cooler than the Kraft break \citep{Ma2021,Weinberg2024}.

We could relax this assumption by grouping systems according to the other
relevant system parameters.
For systems with a given set of parameters $\theta$ (planet mass, host star mass, etc), the number density given by our formal solution is (Equation \ref{eq:formal_solution}):
\begin{equation}
n_\theta (P,t)=\frac{\tau_{\theta,\,\mathrm{mig}}}{P}\int_P^{P_{\theta,t}}S_\theta(P')dP'\,.
\end{equation}
Here, the migration rate and source function, characterized by $\tau_{\theta,\,\mathrm{mig}}$ and $S_\theta$, may depend on the system parameters $\theta$. The total number density is 
\begin{equation}
    n(P,t)=\sum_\theta n_\theta(P,t)\,,
\end{equation}
which can be fitted to the observed occurrence rate.

We also assumed that the tidal migration
law is a power law (Equation \ref{eq:tau_mig}). The realistic tidal decay rate may have a more complicated dependence on the period, such as the case described
in Section \ref{sec:individual_migration_rate} in which temporary resonances cause spurts of faster migration.
Appendix \ref{app:general_decay_rate} shows
that our formal solution (Equation \ref{eq:formal_solution}) still holds under any general form of the orbital decay rate, as long as it remains time-independent, i.e., $\tau_\mathrm{mig}=\tau_\mathrm{mig}(P)$. In the general case, $P_t$ must be understood as the period for which a planet requires time $t$
to migrate to $P$, rather than being
given by Equation \ref{eq:Pt}. Hence, our analysis based on the formal solution can be generalized with an arbitrary parameterization of $\tau_\mathrm{mig}$.

In this work, we did not include a sink function to account for tidal disruption or
engulfment at short orbital periods.
A sink function could be
included on the right side of the continuity equation (Equation \ref{eq:continuity}). As tidal disruption may only happen at ultra-short periods ($P\lesssim 0.5\,\mathrm{days}$), which is beyond the period range of our data, omitting the removal of hot Jupiters may not affect our analysis much. Nevertheless, our analytical framework can be easily generalized if one wants to take related processes into account.

We have only investigated how the occurrence rate depends on orbital period, which leaves some degeneracy between the parameters describing tidal migration ($\chi_\tau$ and $P_\mathrm{mig}$) and the source function ($\chi_S$ and $P_\mathrm{cut}$). However, our general time-dependent solution for the planet number density allows its evolution with age to be investigated, potentially providing a way to break this degeneracy. In particular, a younger sub-population of hot Jupiters would probe $n(P,t)$ at smaller $t$, when the planets have undergone less tidal migration and their period distribution therefore more closely reflects the source function. At present, this approach is limited by the difficulty of obtaining sufficiently accurate ages for hot-Jupiter systems. Future observations, e.g., by finding hot Jupiters in star clusters, or around
stars with detectable asteroseismic oscillations, may provide improved age measurements and enable additional constraints.

In summary, the formalism we developed in this work can be generalized if the above assumptions need to be relaxed. However, such generalization typically introduces additional parameters that would be
challenging to constrain with current observations of the hot Jupiter population. We hope that with the improvement of the data from current and future planetary missions (e.g., PLATO; \citealt{PLATO}), our formalism can be adapted to include more complicated physical processes. The analytical framework we developed may also be helpful for future population synthesis studies of exoplanet demographics.

\section{Conclusion}
\label{sec:conclusion}

It has been proposed that tidal migration might explain the existence of hot Jupiters with orbital periods of days or less. In this picture, the planets excite tides on their host stars after circularization, which dissipate and cause their orbits to decay. 

This should leave traces in the present-day population. In this work, we developed an analytical framework that shows how the current number density of short period hot Jupiters (the observable) can be related to the planetary migration rate and the ``source function'' that adds hot Jupiters to the population. We obtained the general formal solution for planet number density (Equation \ref{eq:formal_solution}), assuming a parameterized migration rate with a power-law dependence on period, i.e., $\tau_\mathrm{mig}\propto P^{\chi_\tau}$. We analyzed this solution for two specific forms of source functions, namely a power-law source on orbital period (Equation \ref{eq:powerlaw_source}) and a Gaussian source function over semi-major axis (Equation \ref{eq:gaussian_source}), and we showed that the migration rate can indeed be constrained, in principle.

We compared the general solutions with a newly measured hot Jupiter occurrence-rate that has been corrected for survey bias. The data favor solutions with a power-law source with $\chi_S<-1$, where $\chi_S$ is the power-law index of the source's dependence over periods. We fitted our numerical solutions to the data, finding three sets of parameters (solutions) with different physical interpretations.

We compared the inferred migration rates from our solutions with the theoretical predictions from the literature, which are mostly characterized by the power-law index of their dependence on periods, $\chi_\tau$, and the characteristic period $P_\mathrm{mig}$ below which the planets have significantly migrated. Solution 1 features $\chi_\tau=3$, which is roughly consistent with the theoretical prediction $\chi_\tau=10/3$ from the ``fast tides suppression'' model \citep{Goldreich1977,Goodman1997}, though the $P_\mathrm{mig}$ predicted by this model might be too short compared to what we inferred from data.
Solution 2, with $\chi_\tau=1.7$, does not
match any model we are aware of.
Solution 3 has $\chi_\tau=5.6$, which is consistent with the ``fast tides enhancement'' model \citep{Terquem2021} and the ``weakly nonlinear wave dissipation'' model \citep{Essick2016}, both of which predict $P_\mathrm{mig}$ values that are compatible with what we inferred.

We discussed the inferred source function of each solution, which describes the injection of circularized hot Jupiters into the population
before they start migrating. Solutions 1 and 2 indicate a strong peak at $\simeq 3.7$ days. This is consistent with the predictions of a high-eccentricity migration model, where rapid circularization by f-mode dissipation only sets in below a certain periastron distance \citep{Wu2018}. Solution 3, on the other hand, peaks at an unexpectedly short period that would likely only be compatible with efficient migration in a proto-planetary disk.

We also compared the inferred migration rates with the rates constrained for individual hot Jupiters, including an updated constraint for WASP-19\,b that we performed in this work (Appendix \ref{app:wasp19b}). None of the models are consistent with the rates obtained for all systems. WASP-12\,b remains the only system with solid evidence of orbital decay, and only solution 3
is (marginally) compatible with the measured decay rate. The strict upper limit of $|\dot{P}|$ for WASP-19\,b favors solution 2.

The migration rates and number density inferred from our solutions predict a rate of planet engulfment or tidal disruption. Recently, \cite{De2023} and \cite{Lau2025} interpreted the subluminous red nova event ZTF-SLRN-2020 as a hot Jupiter engulfed by its host star. We estimated the Galactic rate of such events based on our solutions. We find that solutions 1 and 3 can predict a rate that is roughly consistent with the  rate estimated empirically
by \cite{De2023}.

Although we do not claim that any of our
solutions is definitive,
on balance, solution 1 seems the most appealing. This solution is consistent with (i) an existing tidal migration model (``fast tide suppression''); (ii) the source function expected from high-eccentricity migration; and (iii) the inferred rate of hot Jupiter engulfment from transient observations. The simplest estimates of $P_{\rm mig}$ in this theory are, however, too small compared to our observational inferences. In addition, solution 1 is inconsistent with the migration rates measured for WASP-12\,b and bounded for WASP-19\,b, but there are
reasons why individual measurements might
depart from the population- and time-averaged quantities described by the model. Solution 3 is also attractive but requires a very different physical model, with some processes (e.g., disk migration) ``sourcing'' hot Jupiters to very small periods and tidal dissipation that depends much more strongly on orbital period than in solution 1.

Our inferred values for $\chi_\tau$ disagree with those inferred by \cite{Millholland2025}, who performed a population-level analysis and inferred $0\lesssim \chi_\tau\lesssim 2.33$. We trace this discrepancy to the nonlinear transformation from $P$ to $\tau_\mathrm{d}$,
which maps the number density to a unique $+1$ power law in $\tau_\mathrm{d}$-space for small $\chi_\tau$. By working directly
in period space, our model retains sensitivity
to the migration law in the $\chi_\tau\rightarrow 0$ limit. Our analysis also makes use of
the full time-dependent solution and an updated hot-Jupiter period distribution, although the
new data themselves are not responsible for the
different conclusions.

Our model is deliberately simplified, but the framework can
be generalized as the data improve.
A larger hot-Jupiter census with more diverse stellar hosts,
longer transit-timing baselines,
and new candidate engulfment events from wide-field
variability surveys
will provide increasingly powerful complementary tests.
We hope our model will serve as a useful
analytic framework in the continuing
effort to understand how stars dissipate tides and
how hot Jupiters form and evolve.

\section*{Acknowledgements}

We thank Sarah Millholland, Janosz Dewberry, Jeremy Goodman, Tousif Islam, and Andrew Howard for helpful discussions.  EQ thanks Phil Arras and Nevin Weinberg for early work together on this problem that never saw the light of day.   LM is supported by the Lyman Spitzer, Jr. Postdoctoral Fellowship of Princeton University and the Gordon and Betty Moore Postdoctoral Award of the Kavli Institute for Theoretical Physics.
YS acknowledges support from the Lyman Spitzer, Jr. Postdoctoral Fellowship at
Princeton University and from the Natural Sciences and Engineering Research
Council of Canada (NSERC) [funding reference CITA 490888-16]. The collaborations between the authors have benefited from the Mitchel Postdoctoral Scholar Career Development Fund. This work is dedicated to George Green (1793-1841), whose remarkable life remains less widely known than his profound contributions to mathematics and physics.

\section*{Data Availability}

The source code supporting the analysis and plots within this article is available upon reasonable request to the corresponding author. 

\bibliographystyle{mnras}
\bibliography{bibliography}

\appendix
\onecolumn

\section{Definitions of power-law indices}
\label{sec:definition_of_power_law_indices}

Because several different power laws of period appear in this work, and might be confused with each other,
this Appendix summarizes their definitions. The power-law indices are universally denoted by $\chi$, with subscripts describing the variables.

The tidal migration timescale $\tau_\mathrm{mig}$ scales as follows:
\begin{equation}
    \tau_\mathrm{mig}=-\frac{P}{\dot{P}}=\tau_0\,\bigg(\frac{P}{P_0}\bigg)^{\chi_\tau}\,.
\end{equation}
where $\tau_0$ is the migration timescale at some reference period $P_0$. Introducing the tidal quality factor $Q'$, defined in Equation \ref{eq:tidal_Q}, yields
\begin{equation}
     Q'\propto P^{\chi_\tau-13/3}\,.
\end{equation}
We only consider the case $\chi_\tau\geq0$, since that is a general feature of realistic tidal migration models.

The source function $S(P)$, when assumed to have a power-law dependence on $P$, scales as:
\begin{equation}
     S(P)\propto P^{\chi_S}\,,
\end{equation}
Finally, for the purpose of comparing the power-law index of the $\ln P$-space number density over period with the data, we define
\begin{equation}
    \chi_{\tilde{n}}\equiv\frac{\partial \ln\tilde{n}}{\partial \ln P}\,.
\end{equation}
Since the $\ln P$-space number density, $\tilde{n}$, satisfies $\tilde{n}\,d\ln P=n\,dP$, it follows that $\tilde{n}=nP$.

\section{General Expression of Number Density}
\label{sec:number_density_expression}

In this section, we solve Equation \ref{eq:continuity}. 
Inspired by \cite{Millholland2025},
we define the tidal decay time $\tau_\mathrm{decay}$:
\begin{empheq}[left={\tau_\mathrm{decay}(P)\equiv\empheqlbrace}]{equation}
\label{eq:tau_decay_appendix}
  \begin{aligned}
      &\frac{\tau_0}{\chi_\tau}\bigg[\bigg(\frac{P}{P_0}\bigg)^{\chi_\tau}-1\bigg]&&,\;\chi_\tau>0\,. \\
       &\tau_0\ln(P/P_0)&&,\; \chi_\tau=0\,.
  \end{aligned}
\end{empheq}

By choosing $P_0$ to be less than $P$, one can verify by integrating Equation \ref{eq:tau_mig} that $\tau_\mathrm{decay}(P)$ is the time
required for a planet to migrate from $P$ to 
$P_0$.\footnote{The relationship between M25's $\tau_\mathrm{d}$ and
our $\tau_\mathrm{decay}$ is $\tau_\mathrm{d}=\tau_\mathrm{decay}+\tau_0/\chi_\tau$. Therefore, for $\chi_\tau>0$, $\tau_\mathrm{d}$ is the time required for a planet to migrate to $P=0$. However, for $\chi_\tau=0$, the time to migrate
to $P=0$ diverges, whereas $\tau_\mathrm{decay}$ remains well-defined.} 
Because $\tau_\mathrm{decay}$ is a monotonic function of $P$, this equation can be interpreted as a coordinate transformation from $P$ to $\tau_\mathrm{decay}$. One can show that under this transformation, the continuity equation \ref{eq:continuity} becomes a 1D advection equation with constant velocity $-1$:
\begin{equation}
\label{eq:continuity_tau_appendix}
    \frac{\partial n_\tau}{\partial t}-\frac{\partial n_\tau}{\partial \tau_\mathrm{decay}}=S_\tau\,,
\end{equation}
where $n_\tau=n_\tau(\tau_\mathrm{decay},t)$ and $S_\tau=S_\tau(\tau_\mathrm{decay})$ denote the number density and source function in $\tau_\mathrm{decay}$-space, which are related to their $P$-space definitions by the Jacobian of the transformation:
\begin{equation}
    \label{eq:n_transform}
    n(P,t)=\bigg(\frac{d\tau_\mathrm{decay}}{dP}\bigg)\,
    n_\tau(\tau_\mathrm{decay},t) =
    \frac{\tau_0}{P}\,\bigg(\frac{P}{P_0}\bigg)^{\chi_\tau}n_\tau\,,
\end{equation}
\begin{equation}
\label{eq:S_transform}
S(P) = 
\bigg(\frac{d\tau_\mathrm{decay}}{dP}\bigg)\,
S_\tau(\tau_\mathrm{decay})\,,
\end{equation}
where we make use of Equation \ref{eq:tau_decay_appendix}.

We assume that $n_\tau(\tau_\mathrm{decay},0)=0$. We can then solve Equation \ref{eq:continuity_tau_appendix} with the Green's function method \citep{GreenEssay}. The formal solution is
\begin{equation}
\label{eq:n_tau} n_\tau(\tau_\mathrm{decay},t)=\int^t_0S_\tau(\tau_\mathrm{decay}+t-\tau')d\tau'\,.
\end{equation}
Conceptually, the planet population observed to have a given
value of the coordinate $\tau_\mathrm{decay}$ consists of planets born between $\tau_\mathrm{decay}$ and $\tau_\mathrm{decay}+t$ per unit time. This result is physically expected, because all planets migrate at the same speed in $\tau_\mathrm{decay}$-space and spend equal amounts of time in equal
intervals of $\tau_\mathrm{decay}$.

We can then use Equations \ref{eq:n_transform} and $\ref{eq:S_transform}$ to obtain $n(P,t)$ from Equation \ref{eq:n_tau}, which gives:
\begin{equation}
\label{eq:num_density}
    n(P,t)=\frac{\tau_\mathrm{mig}}{P}\int_P^{P_t}S(P')dP'\,,
\end{equation}
where 
\begin{empheq}[left={P_t=\empheqlbrace}]{equation}
  \begin{aligned}
      & P(1+\chi_\tau t/\tau_\mathrm{mig})^{1/\chi_\tau}&&,\;\chi_\tau>0,\\
      &P\exp(t/\tau_\mathrm{mig})&&,\;\chi_\tau=0\,.
  \end{aligned}
\end{empheq}
The function $P_t(P,t)$ is the period for which a planet requires a time $t$
to migrate to a period $P$,
as can be verified by integrating Equation \ref{eq:tau_mig}.
Therefore, all planets born with $P>P_t$ have not yet had enough time to migrate to $P$, and cannot contribute to the number density at $P$.  This explains why $P_t$ is the upper limit of the integral in Equation \ref{eq:num_density}.

\section{Analysis of different cases}
\label{sec:analysis_details}
In this section, we analyze the different cases as described in Section \ref{sec:solution_analysis}.

\subsection{$\chi_\tau=0$, resonance locking}
\label{sec:analysis_rl}

When $\chi_\tau=0$, the planets migrate as predicted by the theory of resonance locking. In this case, $\tau_\mathrm{mig}(P)=\tau_0$ is independent of period, and the number density is given by
\begin{equation}
    n(P,t)=\frac{\tau_0}{P}\int^{P\exp(t/\tau_0)}_PS(P')\,dP'\,,
\end{equation}
which generally depends on the shape of the source function $S$. Specifically, if $S\propto P^{\chi_S}$ between $P$ and $P_t=P\exp(t/\tau_0)$, then
\begin{empheq}[left={n(P,t)\propto\empheqlbrace}]{equation}
  \begin{aligned}
      & \frac{\tau_0}{P}\frac{P^{\chi_S+1}}{\chi_S+1}(e^{(\chi_S+1)t/\tau_0}-1)\propto P^{\chi_S}\,,&&\chi_S\neq -1\,,\\
      &\frac{t}{P}\propto P^{\chi_S}\,,&&\chi_S= -1\,,
  \end{aligned}
\end{empheq}
i.e., the number density has the same power-law dependence on period as the source function. 

\subsection{$\chi_\tau\sim \mathcal{O}(1)>0$, period-dependent migration}
\label{sec:analysis_freq_dependent_migration}

When $\chi_\tau\sim \mathcal{O}(1)>0$, $\tau_\mathrm{mig}$ is a strong power-law function of the period. It is useful to define the migration period $P_\mathrm{mig}$:
\begin{equation}
P_\mathrm{mig}\equiv P_0\bigg(\frac{\chi_\tau t}{\tau_0}\bigg)^{1/\chi_\tau}\,,  
\end{equation}
such that $\tau_\mathrm{mig}(P_\mathrm{mig})=\chi_\tau t$. Because $\chi_\tau>0$, we have $\tau_\mathrm{mig}(P)\ll\chi_\tau t$ when $P\ll P_\mathrm{mig}$, and $\tau_\mathrm{mig}(P)\gg\chi_\tau t$ when $P\gg P_\mathrm{mig}$. Therefore, $P_\mathrm{mig}$ roughly corresponds to the period below which the tidal migration is considered ``fast'' at the elapsed time $t$. With this definition, we can re-express the number density (Equation \ref{eq:num_density}) as:
\begin{equation}
\label{eq:n_in_P_mig}
    n(P,t)=\frac{\chi_\tau t}{P_\mathrm{mig}}\bigg(\frac{P}{P_\mathrm{mig}}\bigg)^{\chi_\tau -1}\int_P^{P(1+(P/P_\mathrm{mig})^{-\chi_\tau})^{1/\chi_\tau}}S(P')\,dP'\,.
\end{equation}

In the slow-migration limit ($P\gg P_\mathrm{mig}$), we have $(P/P_\mathrm{mig})^{-\chi_\tau}\ll 1$ as $\chi_\tau>0$. This allows us to perform a first-order expansion of the upper limit of the integral in Equation \ref{eq:n_in_P_mig},
reducing the number density to
\begin{equation}
\label{eq:slow_migration}
\begin{split}
    n(P,t)\approx&\;\frac{\chi_\tau t}{P_\mathrm{mig}}\bigg(\frac{P}{P_\mathrm{mig}}\bigg)^{\chi_\tau -1}\int_P^{P(1+(1/\chi_\tau)(P/P_\mathrm{mig})^{-\chi_\tau})}S(P')dP'\\
    \approx&\;\frac{\chi_\tau t}{P_\mathrm{mig}}\bigg(\frac{P}{P_\mathrm{mig}}\bigg)^{\chi_\tau -1}S(P)\bigg[P\bigg(1+\frac{1}{\chi_\tau}\bigg(\frac{P}{P_\mathrm{mig}}\bigg)^{-\chi_\tau}\bigg)-P\bigg]\\
    =&\;tS(P)\,,
\end{split}
\end{equation}
i.e., the number density is simply the time-integrated source function.

In the fast-migration limit ($P\ll P_\mathrm{mig}$), the number density generally depends on both the source function $S$ and the tidal migration law, characterized by $\chi_\tau$. As explained in the main text, it is useful to examine the $\ln P$-space number density $\tilde{n}$, which is characterized by the power-law index 
\begin{equation}
    \chi_{\tilde{n}}=\chi_n+1\equiv\frac{\partial \ln n}{\partial \ln P}+1\,.
\end{equation}

From Equation \ref{eq:n_in_P_mig}, when $n\neq 0$, we have $\chi_n=\chi_\tau-1+\chi_I$, where
\begin{equation}
    \chi_\mathrm{I}\equiv\frac{\partial \ln I(P,t)}{\partial \ln P},\;\mathrm{and}\; I(P,t)=\int^{P_t}_PS(P')dP'\,.
\end{equation}
Therefore $\chi_{\tilde{n}}=\chi_\tau+\chi_I$. For convenience, we can express $\chi_I$ in the following alternative form:
\begin{equation}
\label{eq:chi_I}
    \chi_\mathrm{I} = \frac{P}{I}\frac{\partial I}{\partial P}=\frac{P_t}{I}\frac{\partial \ln P_t}{\partial \ln P}\frac{\partial }{\partial P_t}\bigg(\int^{P_t}_0S(P')dP'\bigg)-\frac{P}{I}\frac{\partial }{\partial P}\bigg(\int^{P}_0S(P')dP'\bigg)=\frac{1}{I}\bigg(P_tS(P_t)\frac{\partial \ln P_t}{\partial \ln P}-PS(P)\bigg)\,.
\end{equation}
We analyze $\chi_I$ for two different choices of the
functional form of $S(P)$.

\subsubsection{Truncated power-law source function}
\label{appendix:truncated_power_law}
In the first case, we assume that $S(P)$ is a power-law with
index $\chi_S$ and a lower truncation at $P_\mathrm{cut}$:
\begin{empheq}[left={S(P)=\empheqlbrace}]{equation}
  \begin{aligned}
      & 0&&,\; P<P_\mathrm{cut},\\
      &S_0\bigg(\frac{P}{P_\mathrm{cut}}\bigg)^{\chi_S}&&,\; P\geq P_\mathrm{cut}\,.
  \end{aligned}
\end{empheq}
 
In the fast-migration limit ($P\ll P_\mathrm{mig}$), we have:
\begin{equation}
    P_t=P\left[1+\left(\frac{P}{P_\mathrm{mig}}\right)^{\!-\chi_\tau}\right]^{1/\chi_\tau}=P_\mathrm{mig}\left[1+\left(\frac{P}{P_\mathrm{mig}}\right)^{\!\chi_\tau\,}\right]^{1/\chi_\tau}\approx P_\mathrm{mig}\,.
\end{equation}
This means when $P_\mathrm{mig}/P_\mathrm{cut}\ll 1$, we have $P_t/P_\mathrm{cut}<1$, which makes $S(P)=0$ between $P$ and $P_t$. Therefore, $n(P,t)=0$ according to the general solution for number density \ref{eq:num_density}. This is an expected result as it corresponds to the scenario where the innermost possible planet
delivered by the source function has not yet migrated to period $P$; hence, no planets can exist with period $P$.

When $P_\mathrm{mig}/P_\mathrm{cut}\gtrsim 1$, there will be two cases depending on the period $P$. When $P<P_\mathrm{cut}\lesssim P_\mathrm{mig}<P_t$,
\begin{empheq}[left={I(P,t)=\empheqlbrace}]{equation}
  \begin{aligned}
      & S_0 P_\mathrm{cut}\ln(P_t/P_\mathrm{cut})&&,\;\chi_S=-1,\\
      &\frac{S_0 P_\mathrm{cut}}{\chi_S+1}\bigg[\bigg(\frac{P_t}{P_\mathrm{cut}}\bigg)^{\chi_S+1}-1\bigg]&&,\; \chi_S\neq-1\,.
  \end{aligned}
\end{empheq}
We can normalize $P$ and $P_t$ to $P_\mathrm{mig}$ and calculate $\chi_I$ using the above expression and Equation \ref{eq:chi_I}:
\begin{empheq}[left={\chi_\mathrm{I}=\empheqlbrace}]{equation}
\label{eq:chi_I_P_ll_Pmin}
  \begin{aligned}
      & \frac{\partial \ln P_t/\partial \ln P}{\ln(P_t/P_\mathrm{mig})+\ln(P_\mathrm{mig}/P_\mathrm{cut})}&&,\;\chi_S=-1,\\
      &(\chi_S+1)\frac{(P_t/P_\mathrm{mig})^{\chi_S+1}(\partial \ln P_t/\partial \ln P)}{(P_t/P_\mathrm{mig})^{\chi_S+1}-(P_\mathrm{cut}/P_\mathrm{mig})^{\chi_S+1}}&&,\; \chi_S\neq-1\,.
  \end{aligned}
\end{empheq}

Because $(P/P_\mathrm{mig})^{\chi_\tau}\ll 1$ in the fast-migration limit, we can write the following terms in Equation \ref{eq:chi_I_P_ll_Pmin} as series expansions in the parameter $(P/P_\mathrm{mig})^{\chi_\tau}$:
\begin{equation}
\label{eq:expansion1}
    \bigg(\frac{P_t}{P_\mathrm{mig}}\bigg)^{\chi_S+1}=\bigg(1+\bigg(\frac{P}{P_\mathrm{mig}}\bigg)^{\chi_\tau}\bigg)^\frac{\chi_S+1}{\chi_\tau}=1+\sum_{k=1}^\infty\binom{\frac{\chi_S+1}{\chi_\tau}}{k}\bigg(\frac{P}{P_\mathrm{mig}}\bigg)^{k\chi_\tau}\,,
\end{equation}
\begin{equation}
\label{eq:expansion2}
    \ln\bigg(\frac{P_t}{P_\mathrm{mig}}\bigg)=\frac{1}{\chi_\tau}\ln\bigg(1+\bigg(\frac{P}{P_\mathrm{mig}}\bigg)^{\chi_\tau}\bigg)=\frac{1}{\chi_\tau}\sum_{k=1}^\infty\frac{(-1)^{k-1}}{k}\bigg(\frac{P}{P_\mathrm{mig}}\bigg)^{k\chi_\tau}\,,
\end{equation}
\begin{equation}
\label{eq:expansion3}
    \frac{\partial \ln P_t}{\partial \ln P}=\bigg(\frac{P}{P_\mathrm{mig}}\bigg)^{\chi_\tau}\bigg(1+\bigg(\frac{P}{P_\mathrm{mig}}\bigg)^{\chi_\tau}\bigg)^{-1}=\sum_{k=1}^\infty(-1)^{k-1}\bigg(\frac{P}{P_\mathrm{mig}}\bigg)^{k\chi_\tau}\,,
\end{equation}
where
\begin{equation}
    \binom{\frac{\chi_S+1}{\chi_\tau}}{k}=\frac{\left(\frac{\chi_S+1}{\chi_\tau}\right)\left(\frac{\chi_S+1}{\chi_\tau}-1\right)\left(\frac{\chi_S+1}{\chi_\tau}-2\right)\dotsm\left(\frac{\chi_S+1}{\chi_\tau}-k+1\right)}{k\,!}
\end{equation}
is the generalized binomial coefficient.

We substitute the above expansions into Equation \ref{eq:chi_I_P_ll_Pmin}. When $\chi_S=-1$,
\begin{equation}
\label{eq:chi_I_general_when_chi_S_eq_-1}
    \chi_I=\frac{\sum_{k=1}^\infty(-1)^{k-1}\left(\frac{P}{P_\mathrm{mig}}\right)^{k\chi_\tau}}{\frac{1}{\chi_\tau}\sum_{k=1}^\infty\frac{(-1)^{k-1}}{k}\left(\frac{P}{P_\mathrm{mig}}\right)^{k\chi_\tau}+\ln\left(\frac{P_\mathrm{mig}}{P_\mathrm{cut}}\right)}\,.
\end{equation}
Because $P_\mathrm{cut}$ and $P_\mathrm{mig}$ are generally determined by different physics, there is no reason to expect them to closely coincide.
This means $\ln(P_\mathrm{mig}/P_\mathrm{cut})$ should be at least of order unity, and the numerator of the above equation consists only of high-order terms of $(P/P_\mathrm{mig})^{\chi_\tau}\ll 1$. Therefore $\chi_I\ll \chi_\tau\sim \mathcal{O}(1)$. Hence, we have $\chi_{\tilde{n}}=\chi_\tau+\chi_I\approx\chi_\tau$.

When $\chi_S\neq -1$, we have
\begin{equation}
\label{eq:chi_I_general_when_chi_S_neq_-1}
    \chi_I=(\chi_S+1)\frac{\left(1+\sum_{k=1}^\infty\binom{\frac{\chi_S+1}{\chi_\tau}}{k}\left(\frac{P}{P_\mathrm{mig}}\right)^{k\chi_\tau}\right)\times\sum_{k=1}^\infty(-1)^{k-1}\left(\frac{P}{P_\mathrm{mig}}\right)^{k\chi_\tau}}{1-\left(\frac{P_\mathrm{cut}}{P_\mathrm{mig}}\right)^{\chi_S+1}+\sum_{k=1}^\infty\binom{\frac{\chi_S+1}{\chi_\tau}}{k}\left(\frac{P}{P_\mathrm{mig}}\right)^{k\chi_\tau}}\,.
\end{equation}
When $P_\mathrm{cut}$ and $P_\mathrm{mig}$ are different, we again expect $|1-(P_\mathrm{cut}/P_\mathrm{mig})^{\chi_S+1}|$ to be at least of order unity, and the numerator of the above equation consists only of high-order terms of $(P/P_\mathrm{mig})^{\chi_\tau}\ll 1$. Therefore $|\chi_I|\ll \chi_\tau\sim \mathcal{O}(1)$. Hence, we still have $\chi_{\tilde{n}}=\chi_\tau+\chi_I\approx\chi_\tau$.

The preceding results are not valid if it happens
to be the case that
$P_\mathrm{cut}\approx P_\mathrm{mig}$. In that case, either $\ln(P_\mathrm{mig}/P_\mathrm{cut})$ or $1-(P_\mathrm{cut}/P_\mathrm{mig})^{\chi_S+1}$ in the denominators of Equation \ref{eq:chi_I_general_when_chi_S_eq_-1} or \ref{eq:chi_I_general_when_chi_S_neq_-1} can become infinitesimal, giving a much larger and potentially period-dependent value of $\chi_I$. This fine-tuned solution is depicted with the
dark brown curves in the numerical solutions shown in Figure \ref{fig:power-law-s}. Physically, in this case, only a small number of planets with periods
just above the $P_\mathrm{cut}$ have migrated to lower periods, causing the number density to fall very steeply 
with decreasing period below $P_\mathrm{cut}\approx P_\mathrm{mig}$.

Another case of interest is when $P_\mathrm{cut}<P<P_\mathrm{mig}<P_t$. In this case, $I(P,t)$ is given by
\begin{empheq}[left={I(P,t)=\empheqlbrace}]{equation}
  \begin{aligned}
      & S_0 P_\mathrm{cut}\ln(P_t/P)&&,\;\chi_S=-1,\\
      &\frac{S_0 P_\mathrm{cut}}{\chi_S+1}\bigg[\bigg(\frac{P_t}{P_\mathrm{cut}}\bigg)^{\chi_S+1}-\bigg(\frac{P}{P_\mathrm{cut}}\bigg)^{\chi_S+1}\bigg]&&,\; \chi_S\neq-1\,.
  \end{aligned}
\end{empheq}
Using Equation \ref{eq:chi_I}, this gives
\begin{empheq}[left={\chi_\mathrm{I}=\empheqlbrace}]{equation}
\label{eq:chi_I_P_gg_Pmin}
  \begin{aligned}
      & \frac{\partial \ln P_t/\partial \ln P-1}{\ln(P_t/P_\mathrm{mig})+\ln(P_\mathrm{mig}/P)}&&,\;\chi_S=-1,\\
      &(\chi_S+1)\frac{(P_t/P_\mathrm{mig})^{\chi_S+1}(\partial \ln P_t/\partial \ln P)-(P/P_\mathrm{mig})^{\chi_S+1}}{(P_t/P_\mathrm{mig})^{\chi_S+1}-(P/P_\mathrm{mig})^{\chi_S+1}}&&,\; \chi_S\neq-1\,.
  \end{aligned}
\end{empheq}

We again use the series expansions \ref{eq:expansion1}--\ref{eq:expansion3} to express $\chi_I$ as:
\begin{empheq}[left={\chi_\mathrm{I}=\empheqlbrace}]{equation}
\label{eq:chi_I_P_gg_Pmin_series}
  \begin{aligned}
      & \frac{\sum_{k=1}^\infty(-1)^{k-1}\left(\frac{P}{P_\mathrm{mig}}\right)^{k\chi_\tau}-1}{\frac{1}{\chi_\tau}\sum_{k=1}^\infty\frac{(-1)^{k-1}}{k}\left(\frac{P}{P_\mathrm{mig}}\right)^{k\chi_\tau}+\ln(P_\mathrm{mig}/P)}&&,\;\chi_S=-1,\\
      &(\chi_S+1)\frac{\left(1+\sum_{k=1}^\infty\binom{\frac{\chi_S+1}{\chi_\tau}}{k}\left(\frac{P}{P_\mathrm{mig}}\right)^{k\chi_\tau}\right)\times\sum_{k=1}^\infty(-1)^{k-1}\left(\frac{P}{P_\mathrm{mig}}\right)^{k\chi_\tau}-\left(\frac{P}{P_\mathrm{mig}}\right)^{\chi_S+1}}{1-\left(\frac{P}{P_\mathrm{mig}}\right)^{\chi_S+1}+\sum_{k=1}^\infty\binom{\frac{\chi_S+1}{\chi_\tau}}{k}\left(\frac{P}{P_\mathrm{mig}}\right)^{k\chi_\tau}}&&,\; \chi_S\neq-1\,.
  \end{aligned}
\end{empheq}

When $\chi_S=-1$, we keep the zeroth-order
leading terms in the numerator and the denominator, giving:
\begin{equation}
    |\chi_I|\approx-\frac{1}{\ln(P/P_\mathrm{mig})}\ll\chi_\tau\sim \mathcal{O}(1)\,.
\end{equation}
Therefore, $\chi_{\tilde{n}}=\chi_\tau+\chi_I\approx\chi_\tau$.

When $\chi_S>-1$, the first-order term
in the numerator of Equation \ref{eq:chi_I_P_gg_Pmin_series} would be $(P/P_\mathrm{mig})^{\chi_\tau}$ or $-(P/P_\mathrm{mig})^{\chi_S+1}$, which are both much smaller than the leading order ($=1$) in the denominator. As we do not expect $\chi_S+1$ to be greater than the order of unity, which would cause the source function to be strongly peaked at long periods, we have:
\begin{equation}
    |\chi_I|\ll|\chi_S+1|\sim \mathcal{O}(1)\,,
\end{equation}
which still indicates $\chi_{\tilde{n}}=\chi_\tau+\chi_I\approx\chi_\tau$.

When $\chi_S<-1$, the leading order in both the numerator and the denominator in Equation \ref{eq:chi_I_P_gg_Pmin_series} would be $(P/P_\mathrm{mig})^{\chi_S+1}\gg 1$. Therefore,
\begin{equation}
    \chi_I\approx\chi_S+1\,,
\end{equation}
which indicates $\chi_{\tilde{n}}=\chi_\tau+\chi_I\approx\chi_\tau+\chi_S+1$.

\begin{figure*}
    \centering
\includegraphics[width=\linewidth]{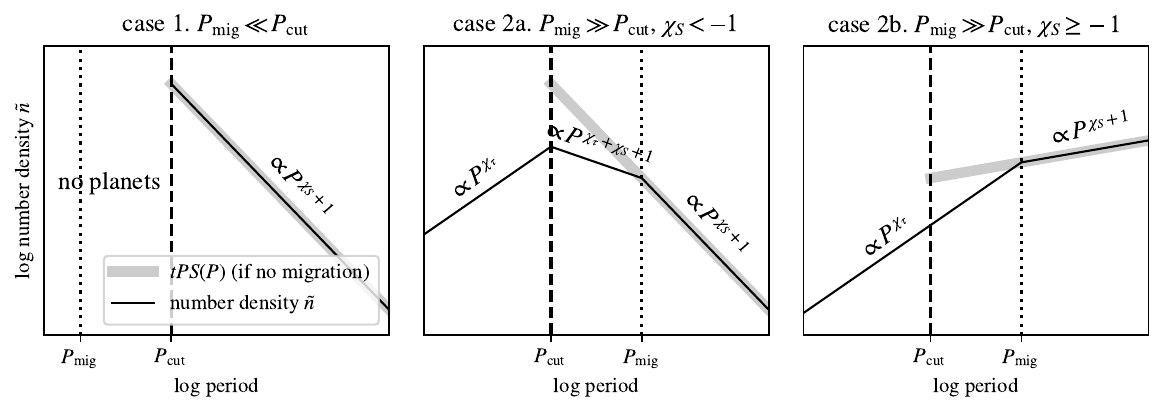}
    \caption{Illustrations of the $\ln P$-space number density $\tilde{n}$ (thin black lines) assuming a source function
    $S\propto P^{\chi_S}$ for $P\geq P_\mathrm{cut}$ and zero for shorter periods. For comparison, a no-migration model is shown, in which $\tilde{n}=tPS(P)$ (thick gray lines). There are a few different cases depending on the relationship between $P_\mathrm{cut}$ and $P_\mathrm{mig}$. 
    \textbf{Case 1}:  $P_\mathrm{mig}\ll P_\mathrm{cut}$. No planets have migrated significantly, and $\tilde{n}\propto P^{\chi_S+1}$ for $P>P_\mathrm{cut}$, resembling $tPS(P)$.
    \textbf{Case 2a}: $P_\mathrm{mig}\gg P_\mathrm{cut}$ and $\chi_S<-1$ (second panel). Planets with $P>P_\mathrm{mig}$
    have not migrated much.
    Planets with $P<P_\mathrm{cut}$
    obey $\tilde{n}\propto P^{\chi_\tau}$,
    the result of tidal migration from
    longer periods.
    Planets with
    $P_\mathrm{cut}<P<P_\mathrm{mig}$ 
    are a mixture of newly-arrived and migrated
    planets, and obey
    $\tilde{n}\propto P^{\chi_\tau+\chi_S+1}$.
    \textbf{Case 2b}: $P_\mathrm{mig}\gg P_\mathrm{cut}$ and $\chi_S\geq -1$. Planets with $P>P_\mathrm{mig}$ have not migrated much, and 
    planets with $P<P_\mathrm{mig}$
    are mainly migrated planets
    obeying $\tilde{n}\propto P^{\chi_\tau}$.
    }
    \label{fig:cartoon}
\end{figure*}

When combined with the cases in the slow-migration limit, where $\chi_{\tilde{n}}=\chi_n+1=\chi_S+1$, the results for the truncated power-law source function can be summarized as follows:
\begin{itemize}
    \item When $P_\mathrm{mig}\ll P_\mathrm{cut}$,
    \begin{empheq}[left={\chi_{\tilde{n}}\approx\empheqlbrace}]{equation}
  \begin{aligned}
      & \mathrm{negligible\;planet\;population} &&,\; P<P_\mathrm{cut},\\
      &\chi_S+1&&,\; P>P_\mathrm{cut}\,.
  \end{aligned}
\end{empheq}\\
\item When $P_\mathrm{mig}\gg P_\mathrm{cut}$,
    \begin{empheq}[left={\chi_{\tilde{n}}\approx\empheqlbrace}]{equation}
  \begin{aligned}
      & \chi_\tau &&,\; P<P_\mathrm{cut},\\
      & \chi_\tau &&,\; P_\mathrm{cut}<P<P_\mathrm{mig}\;\mathrm{if}\;\chi_S\geq -1,\\
      & \chi_\tau+\chi_S+1 &&,\; P_\mathrm{cut}<P<P_\mathrm{mig}\;\mathrm{if}\;\chi_S<-1,\\
      &\chi_S+1&&,\; P>P_\mathrm{mig}\,.
  \end{aligned}
\end{empheq}

\end{itemize}

We illustrate these cases in Figure \ref{fig:cartoon}. Our analysis is consistent with the numerical integration of number density with a power-law source function shown in Figure \ref{fig:power-law-s}.

\subsubsection{Gaussian source function, in semi-major axis $a$}
\label{app:gaussian_source}

\begin{figure}
    \centering
    \includegraphics[width=\linewidth]{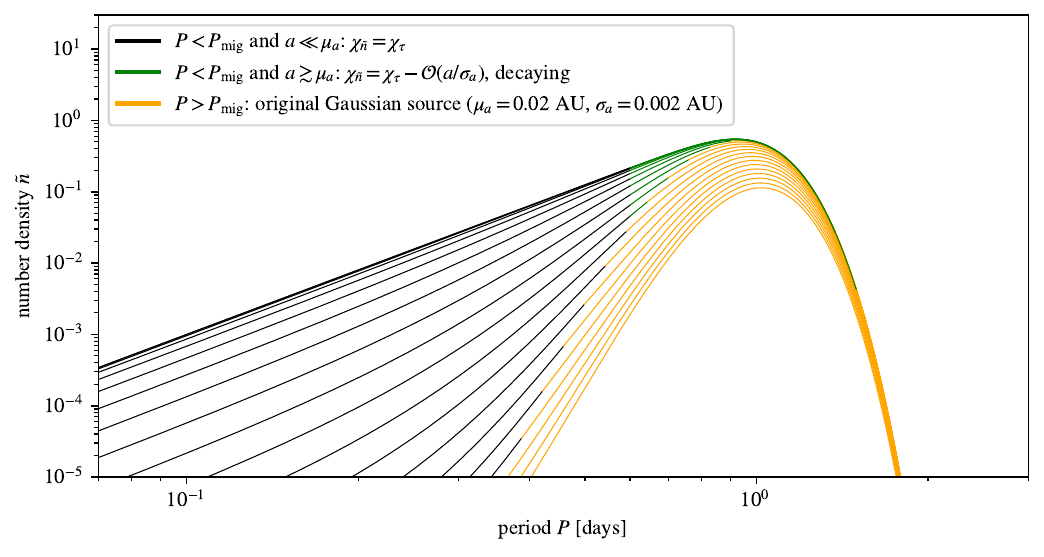}
    \caption{Numerical solutions for the relative $\ln P$-space number density $\tilde n=Pn(P,t)$, assuming $S$ 
    is a Gaussian function of
    semi-major axis $a$ with mean $\mu_a$.
    Different curves are for different
    choices of $P_\mathrm{mig}$. The number density generally increases as time progresses and $P_\mathrm{mig}$ increases. 
    \textbf{Black}: When $P<P_\mathrm{mig}$ and $a < \mu_a$, $\tilde n$ is a power-law with index $\chi_\tau$. 
    \textbf{Green}: As $a$ approaches and surpasses $\mu_a$, the number density starts to decay as $\tilde{n}\propto P^{\chi_\tau-\mathcal{O}(a/\sigma_a)}$. 
    \textbf{Orange}: when $P>P_\mathrm{mig}$, planets migrate slowly and $\tilde n$ resembles the source function in $\ln P$ space.}
    \label{fig:gaussian_source}
\end{figure}

Another physically motivated source function is a Gaussian function of the planetary semi-major axis $a$:
\begin{equation}
    S_a(a)=\frac{S_0}{\sqrt{2\pi \sigma_a^2}}\exp\left(-\frac{(a-\mu_a)^2}{2\sigma_a^2}\right)\,,
\end{equation}
where $\sigma_a$ and $\mu_a$ are the width and mean of the Gaussian. Our usual source function, defined in $P$-space, is related to $S_a(a)$ as:
\begin{equation}
    S(P)=\frac{d a}{dP}S_a(a)=\frac23\frac{a}{P}S_a(a)\,,
\end{equation}
using Kepler's Third Law.

The integral $I(P,t)$ is given by
\begin{equation}
    I(P,t)=\int^{P_t}_PS(P')dP'=\int^{a_t}_aS_a(a')da'=\frac{S_0}{2}\bigg[\mathrm{erf}\bigg(\frac{a_t-\mu_a}{\sqrt{2}\sigma_a}\bigg)-\mathrm{erf}\bigg(\frac{a-\mu_a}{\sqrt{2}\sigma_a}\bigg)\bigg]\,,
\end{equation}
where $a_t\equiv(P_t/P)^{2/3}a$ is the semi-major axis corresponding to the period $P_t$. Using Equation \ref{eq:chi_I}, we have
\begin{equation}
\label{eq:chi_I_gaussian}
    \chi_I=\frac{4}{3\sqrt{\pi}}\frac{{\tilde {a}_t}\exp\left(-\left({\tilde {a}_t}-{\tilde {\mu}_a}\right)^2\right)\left(\frac{\partial \ln P_t}{\partial \ln P}\right)-{\tilde {a}}\exp\left(-\left({\tilde {a}}-{\tilde {\mu}_a}\right)^2\right)}{\mathrm{erf}\left({\tilde {a}_t}-{\tilde {\mu}_a}\right)-\mathrm{erf}\left({\tilde {a}}-{\tilde {\mu}_a}\right)}\,,
\end{equation}
where we defined ${\tilde a_t}\equiv a_t/(\sqrt{2}\sigma_a)$, ${\tilde a}\equiv a/(\sqrt{2}\sigma_a)$, and ${\tilde {\mu}_a}\equiv \mu_a/(\sqrt{2}\sigma_a)$ for convenience. The order of $\chi_I$ will hence depend on the relationship of ${\tilde a}$, ${\tilde a}_t$, and ${\tilde {\mu}_a}$.

In the slow-migration limit, the number density should resemble the Gaussian source function as discussed in Section \ref{sec:analysis_freq_dependent_migration}. Therefore, we only consider the case of fast migration, where $P\ll P_\mathrm{mig}$ and ${\tilde a}_t\gg {\tilde a}$. We further exclude the cases where ${\tilde a}\ll {\tilde a}_t\ll{\tilde {\mu}_a}$ or ${\tilde {\mu}_a}\ll{\tilde a}\ll {\tilde a}_t$ because, in those instances, the only contributions
to the number density are from the exponentially decaying tail of the Gaussian source function, making it practically impossible to constrain due to the very limited number of planets at those distances. Therefore, the only cases of interest are ${\tilde a}\ll{\tilde a}_t\sim{\tilde {\mu}_a}$, ${\tilde a}\sim{\tilde {\mu}_a}\ll{\tilde a}_t$, or ${\tilde a}\ll {\tilde {\mu}_a}\ll {\tilde a}_t$. For the denominator of \ref{eq:chi_I_gaussian}, we have
\begin{empheq}[left={\mathrm{erf}\left({\tilde {a}_t}-{\tilde {\mu}_a}\right)-\mathrm{erf}\left({\tilde {a}}-{\tilde {\mu}_a}\right)\approx\empheqlbrace}]{equation}
  \begin{aligned}
      & \mathrm{erf}\left({\tilde {a}_t}-{\tilde {\mu}_a}\right)-(-1)\sim \mathcal{O}(1)&&,\; {\tilde a}\ll {\tilde a_t}\sim{\tilde {\mu}_a},\\
      & 1-(-1)\sim \mathcal{O}(1)&&,\;{\tilde a}\ll {\tilde {\mu}_a}\ll{\tilde a}_t\\
      &1-\mathrm{erf}\left({\tilde {a}}-{\tilde {\mu}_a}\right)\sim \mathcal{O}(1)&&,\; {\tilde a}\sim{\tilde {\mu}_a}\ll{\tilde a}_t\,.
  \end{aligned}
\end{empheq}
Therefore, the order of $\chi_\mathrm{I}$ is determined by the order of the numerator of Equation \ref{eq:chi_I_gaussian}. When the Gaussian is not too narrow, both ${\tilde {a}}$ and ${\tilde {a}_t}$ should be at most of order unity. Given that $\partial \ln P_t/\partial \ln P\ll 1$ in the fast migration limit, this means that the only way to make $\chi_I$ grow to a
value of order unity is to require ${\tilde {a}}\sim{\tilde \mu_{a}}$, which indicates $\chi_I\sim-\mathcal{O}({\tilde {a}})$ and $\chi_{\tilde n}\sim \chi_\tau-\mathcal{O}(\tilde a)$. In all other cases, we should have $\chi_{\tilde n}\sim \chi_\tau$.

In summary, for a Gaussian source function in $a$, the migrated number density $\tilde n$ depends on period. When $P\ll P_\mathrm{mig}$ and $a\ll \mu_a$, we have ${\tilde n}\propto P^{\chi_\tau}$. As periods become longer, if the semi-major axis $a$ reaches the peak of the Gaussian $\mu_a$ before $P$ reaches $P_\mathrm{mig}$, the number density starts to decay as ${\tilde n}\propto P^{\chi_\tau-\mathcal{O}({\tilde a})}$. When $P$ reaches $P_\mathrm{mig}$, we are again in the slow-migration limit and ${\tilde n}\approx tPS(P)$, resembling the Gaussian source function in $P$-space. Figure \ref{fig:gaussian_source} shows some numerical solutions for the migrated number density, for different choices of $P_\mathrm{mig}$ and a fixed Gaussian source function with $\mu_a=0.02\,\mathrm{AU}$ and $\sigma_a=0.002\,\mathrm{AU}$. The results are consistent with our analysis.

\section{Number Density with A General Orbital Decay Rate}
\label{app:general_decay_rate}

Our formal solution for the planet number density does not depend on our specific assumption regarding the power-law form of the tidal migration rate (Equation \ref{eq:tau_mig}), as long as the rate has no explicit dependence
on time, i.e., $\tau_\mathrm{mig}=\tau_\mathrm{mig}(P)$. To show that, we note that for orbital decay, we have $\dot{P}<0$. Therefore, the number density (Equation \ref{eq:formal_solution}) can be written as:
\begin{equation}
\label{eq:D1}
n(P,t)=-\frac{1}{\dot{P}(P)}\int_P^{P_t(P,t)}S(P')\,dP'\,.
\end{equation}
Here, $P_t$ is redefined as the initial
period of a planet that will migrate
to $P$ after time $t$, and it is generally a function of both $P$ and $t$. This is a generalization of the explicit definition in Equation \ref{eq:Pt}, which assumes that $\dot{P}\propto P^{1-\chi_\tau}$. We can make use of $\dot{P}=dP/dt$ to write down its implicit expression based on this definition:
\begin{equation}
    \label{eq:Pt_general}
    t=\int_{0}^tdt'=\int_{P_t}^P\frac{dP'}{\dot{P}(P')}\,.
\end{equation}

Under this definition, $P_t=P$ when $t=0$, which means \ref{eq:D1} satisfies the initial condition $n(P,0)=0$. To verify that it also satisfies the continuity equation, we substitute \ref{eq:D1} into the left-hand side of Equation \ref{eq:continuity}:
\begin{equation}
\label{eq:D3}
    \frac{\partial n(P,t)}{\partial t}+\frac{\partial(\dot{P}n(P,t))}{\partial P}=-\frac{1}{\dot{P}(P)}\frac{\partial P_t}{\partial t}S(P_t)-\left(\frac{\partial P_t}{\partial P}S(P_t)-S(P)\right)=-S(P_t)\left(\frac{1}{\dot{P}(P)}\frac{\partial P_t}{\partial t}+\frac{\partial P_t}{\partial P}\right)+S(P)\,.
\end{equation}
To obtain $\partial P_t/\partial t$ and $\partial P_t/\partial P$, we take the partial derivatives with respect to $t$ and $P$ on both sides of Equation \ref{eq:Pt_general}, which gives:
\begin{equation}
\label{eq:partial_pt_partial_t}
    1=-\frac{\partial P_t}{\partial t}\frac{1}{\dot{P}(P_t)}\,;
\end{equation}
\begin{equation}
\label{eq:partial_pt_partial_p}
    0=\frac{1}{\dot{P}(P)}-\frac{\partial P_t}{\partial P}\frac{1}{\dot{P}(P_t)}\,.
\end{equation}
As $\dot{P}\neq0$, Equations \ref{eq:partial_pt_partial_t} and \ref{eq:partial_pt_partial_p} can be combined to show that
\begin{equation}
    \frac{1}{\dot{P}(P)}\frac{\partial P_t}{\partial t}+\frac{\partial P_t}{\partial P}=0\,.
\end{equation}
Therefore, the right hand side of Equation \ref{eq:D3} equals $S(P)$. Thus, the number density satisfies the continuity equation and retains its general formal solution, without assuming any specific forms of $\dot{P}(P)$ or $\tau_\mathrm{mig}(P)$.

\section{New Constraints on WASP-19\,b's Orbital Decay}\label{app:wasp19b}

Shortly after WASP-19\,b's discovery, it was identified as one of the most favorable targets in the search for orbital decay \citep{Hebb2010, ValsecchiRasio2014, Essick2016}. Initially, \citet{Patra2020} reported a statistically significant but 
unconvincing detection of a period decrease, based on transit times drawn from heterogeneous sources. Correcting errors in the transit-timing literature and adding new {\it TESS} transit times ruled out rapid orbital decay \citep{Adams2024}. Since 2024, WASP-19 has been observed in {\it TESS} sectors 89, 90, and 99, and a transit was observed with {\it JWST}'s NIRSpec/Prism instrument as part of General Observer program no.\ 5924 (P.I.\ David K. Sing). We derived an updated constraint on the planet's orbital decay rate using \citet{Adams2024}'s curated list of pre-{\it TESS} literature transit times, combined with times we derived from the system's {\it TESS} and {\it JWST} light curves.

For our transit timing analysis, we used the {\it TESS} Science Processing Operations Center (SPOC) light curves, which were phase-folded to determine the consensus transit parameters, and then used to measure the individual transit times. Uncertainties were estimated with a short Markov Chain Monte Carlo (MCMC) analysis that utilized the \texttt{emcee} code \citep{emcee}. After discarding partial transits, we were left with $207$ {\it TESS} transit times with a median uncertainty of $38$~seconds. To measure a transit time from the {\it JWST} data, we used the bandpass-summed light curve from Allen et al.\ (in prep), which was produced using the reduction strategy in Section~3.1.2 of \citet{Allen2026}. First, we masked the portion of the light curve
affected by a prominent starspot crossing. Then, we performed a joint MCMC analysis with a transit light-curve model and quadratic detrending. The best-fit transit time was $t_\mathrm{mid}\,{=}\,2460757.1041459$\,BJD with an uncertainty of $0.57$~s.

Our $303$-transit dataset spans $17.2$ years and shows no compelling evidence for departures from a constant period. We derived constraints on the planet's decay rate using an MCMC analysis of a model in which the period
is steadily decreasing, parameterized in the same way as the model presented by
\cite{Patra2017}. The resulting $dP/dN$ posterior was very nearly Gaussian. We measured $dP/dt\,{=}\,0.07\,{\pm}\,0.29\,\mathrm{ms/yr}$, implying a $3\sigma$ upper limit on the decay rate of $0.80~\mathrm{ms/yr}$.

\bsp
\label{lastpage}
\end{document}